\documentclass[twocolumn,showpacs,aps,prd]{revtex4-1}

\RequirePackage[colorlinks,breaklinks,pdftitle={ATLAS draft},pdfauthor={The ATLAS Collaboration}]{hyperref}  
\hypersetup{linkcolor=blue,citecolor=blue,filecolor=black,urlcolor=blue} 

\usepackage{graphicx} 
\usepackage{atlasphysics} 
\usepackage{rotating} 
\usepackage{multirow}

\newcommand{\Etiso}   {\ensuremath{E_{\mathrm{T}}^{\mathrm{iso}}}}

\usepackage{preprintcover}  
\PreprintCoverPaperTitle{Measurement of the production cross section of an isolated photon associated 
  with jets in proton-proton collisions at $\sqrt{s}= 7\TeV$ with the ATLAS detector}
\PreprintIdNumber{CERN-PH-EP-2012-009}  
\PreprintCoverAbstract{A measurement of the cross section for the
  production of an isolated photon in association with jets in proton-proton
  collisions at a center-of-mass energy $\sqrt{s} = 7\TeV$ is presented.
  Photons are reconstructed in the pseudorapidity range 
  $|\eta^\gamma|<1.37$ and with a transverse energy $\ET^{\gamma}> 25 \GeV$. 
  Jets are reconstructed in the rapidity range $|y^{\rm jet}|<4.4$ and with a
  transverse momentum $\pt^{\rm jet} > 20 \GeV$. 
  The differential cross section $d\sigma/d\ET^\gamma$ is measured,
  as a function of the photon transverse energy, for three
  different rapidity ranges of the leading-$\pt$ jet: $|y^{\rm jet}| < 1.2$,
  $1.2 \leq |y^{\rm jet}| < 2.8$ and $ 2.8 \leq |y^{\rm jet}| < 4.4$.
  For each rapidity configuration the same-sign $(\eta^\gamma y^{\rm jet}\ge 0)$
  and opposite-sign $(\eta^\gamma y^{\rm jet}< 0)$ cases are studied separately.  
  The results are based on an integrated luminosity of 37~pb$^{-1}$, collected with
  the ATLAS detector at the LHC. 
  Next-to-leading order perturbative QCD calculations are found to be in fair
  agreement with the data, except for $\ET^{\gamma}< 45$~GeV, where
  the theoretical predictions overestimate the measured cross sections.}
\PreprintJournalName{Physical Review D}  

\begin{document}

\vspace{0.5cm}

\title{\boldmath Measurement of the production cross section of an isolated photon associated 
  with jets in proton-proton collisions at $\sqrt{s}= 7\TeV$ with the ATLAS detector}

\author{G.\ Aad \textit{et al.}}\thanks{Full author list given at the end of the article.} 
\collaboration{ATLAS Collaboration}

\date{\today}

\begin{abstract}
  A measurement of the cross section for the
  production of an isolated photon in association with jets in proton-proton
  collisions at a center-of-mass energy $\sqrt{s} = 7\TeV$ is presented.
  Photons are reconstructed in the pseudorapidity range 
  $|\eta^\gamma|<1.37$ and with a transverse energy $\ET^{\gamma}> 25 \GeV$. 
  Jets are reconstructed in the rapidity range $|y^{\rm jet}|<4.4$ and with a
  transverse momentum $\pt^{\rm jet} > 20 \GeV$. 
  The differential cross section $d\sigma/d\ET^\gamma$ is measured,
  as a function of the photon transverse energy, for three
  different rapidity ranges of the leading-$\pt$ jet: $|y^{\rm jet}| < 1.2$,
  $1.2 \leq |y^{\rm jet}| < 2.8$ and $ 2.8 \leq |y^{\rm jet}| < 4.4$.
  For each rapidity configuration the same-sign $(\eta^\gamma y^{\rm jet}\ge 0)$
  and opposite-sign $(\eta^\gamma y^{\rm jet}< 0)$ cases are studied separately.  
  The results are based on an integrated luminosity of 37~\ipb, collected with
  the ATLAS detector at the LHC. 
  Next-to-leading order perturbative QCD calculations are found to be in fair
  agreement with the data, except for $\ET^{\gamma}\lesssim 45 \GeV$, where
  the theoretical predictions overestimate the measured cross sections.
\end{abstract}


\maketitle

\section{Introduction}
\label{sec:Introduction}
At colliders, prompt photons are defined as photons produced in the 
beam particle collisions and not originating from particle decays.
They include both \emph{direct} photons, which originate from the hard
process, and \emph{fragmentation} photons, which arise from
the fragmentation of a colored high-\pt\ parton~\cite{jetphox1,jetphox2}.
At the LHC, the production of prompt photons in association with jets in
proton-proton collisions, $pp \to \gamma + {\rm
jet} + X$, 
represents an important test of perturbative QCD predictions
at large hard-scattering scales ($Q^2$) and over a wide range
of the parton momentum fraction ($x$).
In addition the study of the angular correlations between the photon and
the jet can be used to constrain the photon fragmentation functions~\cite{Belghobsi:2009hx}.
Since the dominant $\gamma + {\rm jet}$ production mechanism in $pp$
collisions at the LHC is through the $qg~\rightarrow~q\gamma$ process,
the measurement of the photon + jet cross section 
at high rapidities and low transverse momenta
can also be
exploited to constrain the gluon density function inside the
proton~\cite{Belghobsi:2009hx,PhysRevD.35.1584,promptphoton_and_gluon_pdf,dEnterria:2012yj}
for values of the incoming parton momentum fraction $x$ down to
$\approx \mathcal{O}(10^{-3})$. 
For the same reason, this final state can be used to obtain a
high purity sample of quark-originated jets~\cite{Gallicchio:2011xc}
that can be exploited to study detector performance with respect to
these jets. The same events can also be used to calibrate the jet
energy scale
by profiting from momentum conservation in the transverse plane
and the accurate energy measurement of the photon in
the electromagnetic calorimeter~\cite{JES_gamjet_ConfNote}.
Finally, $\gamma$ + jet events provide one of the main backgrounds
in searches of Higgs bosons decaying to a photon
pair~\cite{higgspaper_2010}.
An accurate knowledge of the photon + jet rate and angular distribution 
can be useful to understand the background level and shape
in these searches.

In this article a measurement of the production cross section
of an isolated prompt photon in association with jets, in $pp$
collisions at a center-of-mass energy $\sqrt{s} = 7\TeV$, is 
presented.
Photons are reconstructed in the pseudorapidity range of
$|\eta^\gamma|<1.37$ and in the transverse energy range of $\ET^{\gamma}>
25\GeV$. The same isolation criterion as used in our
measurements of the inclusive isolated prompt 
photon~\cite{promptphotonpaper_2010, promptphoton_confnote_2011} 
and diphoton production cross sections~\cite{diphoton_2011} is used.
It is based on the amount $\Etiso$ of
transverse energy deposited in the calorimeters inside a cone of
radius $R = \sqrt{\left( \eta - \eta^{\gamma} \right)^{2}
+ \left( \phi - \phi^{\gamma} \right)^2} = 0.4$ centered around the
photon direction (defined by $\eta^\gamma$, $\phi^\gamma$)
\footnote{The ATLAS reference system is a Cartesian
right-handed coordinate system, with the nominal collision point at
the origin. The anticlockwise beam
direction defines the positive $z$-axis, while the positive $x$-axis
is defined as pointing from the collision point to the center of the
LHC ring and the positive $y$-axis points upwards. The azimuthal angle
$\phi$ is measured around the beam axis, and the polar angle $\theta$
is measured with respect to the $z$-axis. Pseudorapidity is defined as
$\eta = -\ln\tan(\theta/2)$, and transverse energy is defined as
$\ET = E\sin\theta$.}. 
The contribution from electromagnetic calorimeter cells in the $(\Delta
\eta, \Delta \phi) = (\pm 0.0625 , \pm 0.0875)$ region
around the photon barycenter is not included in the sum. 
The mean value of the small leakage of the photon energy
outside this region, evaluated as a function of the
photon transverse energy, is subtracted from the measured
value of \Etiso.
The typical size of this correction is a few percent of the
photon transverse energy. 
The measured value of \Etiso~is further corrected by
subtracting the estimated contributions from the underlying event 
and additional inelastic $pp$ interactions.
This correction is computed on an event-by-event basis using the method
suggested in Ref.~\cite{Cacciari:area,Cacciari:UE}.
After the isolation requirement is
applied, the relative contribution to the total cross section from
fragmentation photons decreases, though it remains non-negligible
especially at low transverse energies, below 35-40 GeV~\cite{jetphox2}.
The isolation requirement significantly reduces the main background,
which consists of QCD multijet events where one jet typically contains
a $\pi^{0}$ or $\eta$ meson which carries most of the jet energy and
is misidentified as a prompt photon because it
decays into a photon pair.
Jets are reconstructed in the rapidity range of $|y^{\rm jet}|<4.4$ and
transverse momentum range of $\pt^{\rm jet} > 20 \GeV$. 
The minimum separation between the highest $\pt$ ({\em leading}) jet and
the photon in the $\{\eta,\phi\}$ plane is $\Delta R>1.0$.
The leading jet is required to be in either the
{\em central} ($|y^{\rm jet}| < 1.2$), {\em forward} ($1.2 \leq
|y^{\rm jet}| < 2.8$) or {\em very forward} ($2.8 \leq |y^{\rm jet}| <
4.4$) rapidity interval. 

The differential cross section $d\sigma/d\ET^\gamma$ is measured
for each of the three leading jet rapidity categories. 
Measurements are performed separately for the two cases where the
photon pseudorapidity and the leading jet rapidity have same-sign
$(\eta^\gamma y^{\rm jet}\ge 0)$ or opposite-sign $(\eta^\gamma y^{\rm
jet}< 0)$, and the results are compared to next-to-leading order (NLO)
perturbative QCD theoretical predictions.  
Separating the selected phase space into these six different angular
configurations allows the comparison between data and theoretical
predictions in configurations where the relative contribution of the
fragmentation component to the total cross section is different, and 
in different ranges of $x$, which in the leading-order approximation 
is equal to
$x=\frac{\ET^\gamma}{\sqrt{s}}\left(e^{\pm \eta^\gamma}+e^{\pm y^{\rm jet}}\right)$.
The differential cross sections are measured up to $\ET^\gamma = 400$
GeV for the central and forward jet configurations, and up to
$\ET^\gamma = 200$ GeV for the very forward jet configurations.
These measurements cover the region $x\gtrsim 0.001$
and $625$~GeV$^2\leq Q^2 \equiv (\ET^\gamma)^2 \leq 1.6 \times 10^5$
GeV$^2$, thus extending the kinematic reach of previous photon + jet
measurements at
hadron~\cite{Abazov:2008er,Abe:1997ma,Akesson_promptphoton,Alitti:1992kw}  
and
electron-proton~\cite{Aaron:2010uj,Aaron:2007eh,Chekanov:2006un,Chekanov:2004wr}
colliders.

\section{The ATLAS Detector}
\label{sec:Detector}
The ATLAS experiment~\cite{ATLAS_detector} is a multipurpose particle
physics detector with a forward-backward symmetric
cylindrical geometry and nearly $4\pi$ coverage in solid angle.

The inner tracking detector covers the
pseudorapidity range $|\eta| < 2.5$, and consists of a silicon
pixel detector, a silicon microstrip detector, and,
for $|\eta| < 2.0$, a transition radiation tracker.
The inner detector is surrounded by a thin superconducting solenoid 
providing a 2T magnetic field.

The electromagnetic calorimeter is a lead-liquid argon 
sampling calorimeter.
It is divided into a
barrel section, covering the pseudorapidity region $|\eta|< 1.475$,
and two end-cap sections, covering the pseudorapidity regions
$1.375<|\eta|<3.2$. It consists of three longitudinal layers in most of
the pseudorapidity range. 
The first layer, with a thickness between 3 and 5 radiation lengths, is
segmented into high granularity strips in the $\eta$ direction (width
between 0.003 and 0.006 depending on $\eta$, with the exception of the
regions $1.4<|\eta|<1.5$ and $|\eta|>2.4$), sufficient to provide
event-by-event discrimination between single-photon showers and two
overlapping showers coming from a $\pi^0$ decay. The second layer of
the electromagnetic calorimeter, which collects most of the energy
deposited in the calorimeter by the photon shower, has a thickness
around 17 radiation lengths and a cell granularity of $0.025\times0.025$ in
$\eta\times\phi$. A third layer, with
thickness varying between 4 and 15 radiation lengths, collects the tails of
the electromagnetic showers and provides an additional point to
reconstruct the shower barycenter.
In front of the 
calorimeter a thin presampler layer, covering
the pseudorapidity interval $|\eta|<1.8$, is used to correct for
energy loss before the calorimeter. 
The electromagnetic energy scale is measured using $Z\to ee$ events with
an uncertainty better than 1\%~\cite{electronperf_2011}.
The linearity has been found to be close to 1\%.
At low $|\eta|$ the stochastic term is $(9-10)\%/\sqrt{E [{\rm GeV}]}$.
However, it worsens as the amount of material in front of the calorimeter
increases at larger $|\eta|$.   
The constant term is measured to be about 1.2\% in the barrel 
and 1.8\% in the end-cap region up to $|\eta|<2.47$ which is relevant for
this analysis.

A hadronic sampling calorimeter is located outside the electromagnetic
calorimeter. It is made of scintillating tiles and steel in the
barrel section ($|\eta|<1.7$), with depth around 7.4 interaction
lengths, and of two end-caps of copper and liquid argon,
with depth around 9 interaction lengths. Hadronic jets are reconstructed with
an energy scale uncertainty of the order of 2.5\% in the central to 14\% in the
very forward regions~\cite{jes_uncertainty2}.

The muon spectrometer surrounds
the calorimeters. It consists of three large air-core 
superconducting toroid systems, stations of precision tracking 
chambers providing accurate muon tracking over $|\eta| < 2.7$,
and detectors for triggering over $|\eta| < 2.4$.

Events containing photon candidates are selected
by a three-level trigger system.
The first level trigger
(level-1) is hardware based: using a trigger cell granularity
($0.1\times 0.1$ in $\eta\times\phi$) coarser than that of the electromagnetic
calorimeter, it searches for electromagnetic clusters within a fixed
window of size $0.2\times 0.2$ and retains only those whose total
transverse energy in two adjacent trigger cells is above a programmable
threshold. The algorithms of the second and third level 
triggers (collectively referred to as the {\em high-level trigger})
are implemented in software. The
high-level trigger exploits the full granularity and precision of the
calorimeter to refine the level-1 trigger selection, based on improved
energy resolution and detailed information on energy deposition in the
calorimeter cells.

\section{Collision Data and Simulated Samples}
\label{sec:Samples}
\subsection{Collision Data}
The measurements presented here are based on $pp$ collision
data collected at a center-of-mass energy $\sqrt{s}=7$ TeV 
in 2010.
Only events taken in stable beam conditions are considered and the
trigger system, the tracking devices and the calorimeters are also
required to be operational. 
Events are recorded using two single-photon triggers, with
nominal transverse energy thresholds of 20 and 40 GeV.
During the 2010 data-taking, no prescale was applied to the 40 GeV
threshold trigger and the corresponding total integrated luminosity of
the collected sample amounts to $\int{L dt}=37.1$~\ipb~\cite{lumipaper,ATLAS-CONF-2011-011}.
In this measurement, this threshold is used to collect events in which
the photon transverse energy, after reconstruction and calibration, is
greater than 45 GeV.
During the same data-taking period the average prescale of the 20 GeV
threshold trigger was 5.5, leading to a total integrated luminosity of
$(6.7\pm 0.2)$~\ipb. This threshold is used in this measurement to collect
events in which the photon transverse energy is lower than 45 GeV 

The selection criteria applied by the trigger on shower-shape variables
computed from the energy profiles of the showers in the calorimeters 
are looser than the photon identification criteria applied in this measurement.
Minimum-bias events, triggered by two 
sets of scintillation counters located at $z=\pm3.5$~m from the 
collision center, are used to estimate the single-photon trigger 
efficiencies for true prompt photons with pseudorapidity $|\eta^\gamma|<2.37$.
The efficiencies are constant and consistent with 100\% within the uncertainty
(Sec.~\ref{sec:xsection})
 for $\ET^\gamma>43$~GeV and 
$\ET^\gamma>23$~GeV for the 40 GeV and 20 GeV threshold triggers, respectively.

In order to reduce noncollision backgrounds, events are required to
have a reconstructed primary vertex with at least three associated tracks
and consistent with the average beam spot position.
The inefficiency of this requirement is negligible
in true photon + jet events passing the acceptance criteria.
The estimated contribution to the final photon sample
from noncollision backgrounds is less than 0.1\% and is therefore
neglected~\cite{promptphotonpaper_2010,promptphoton_confnote_2011}.

The total number of selected events in data after the trigger, data quality
and primary vertex requirements is
approximately six million.

\subsection{Simulated events}
\label{subsec:simulated_data}

To study the characteristics of signal and background events,
simulated samples are generated using {\tt PYTHIA}
6.423~\cite{pythia}. 
The event generator parameters, including those of the underlying event
model, are set according to the ATLAS
AMBT1 tune~\cite{ATLAS_MC10_tune}, and the detector response is
simulated using the {\tt GEANT4} program~\cite{geant}.
These samples are reconstructed with the same algorithms
used for data. More details on the event generation and simulation 
infrastructure are provided in Ref.~\cite{ATLAS_simulation}. 
For the evaluation of systematic uncertainties related to the choice
of the event 
generator and parton shower model, alternative samples are
generated with {\tt HERWIG} 6.510~\cite{Herwig}.
The {\tt HERWIG} event generation parameters are set 
according to the AUET1 tune~\cite{ATLAS_MC10_tune_Herwig} and the underlying event
is generated using {\tt JIMMY} 4.31~\cite{JIMMY} with 
multiple parton interactions enabled.

The signal sample includes leading order $\gamma$ + jet events from both
$q g \rightarrow q \gamma$ and $q \bar{q} \rightarrow g \gamma$ hard
scattering 
and from quark bremsstrahlung in QCD dijet events.
The background sample is generated by using all tree-level
2$\rightarrow$2 QCD processes, removing $\gamma$ + jet events from quark 
bremsstrahlung.

The ratio between selected diphoton and inclusive photon + jet events is
estimated to be 0.3\% using {\tt PYTHIA} diphoton samples.
Therefore, background from diphoton events is neglected.

\section{Photon and Jet Selection}
\label{sec:Selection}
\subsection{Photon selection}
\label{subsec:photon_selection}
Photons are reconstructed starting from clusters in the electromagnetic
calorimeter with transverse energies exceeding 2.5~GeV, measured in
projective towers of 3$\times$5 cells in $\eta\times\phi$ in the
second layer of the calorimeter. An attempt is made to match these
clusters with tracks that are reconstructed in the inner detector and
extrapolated to the calorimeter. Clusters without matching tracks are
classified as {\em unconverted} photon candidates. Clusters
with matched tracks are classified as electron candidates. To recover
photon conversions, clusters matched to pairs of tracks originating
from reconstructed conversion vertices in the inner detector or to single
tracks with no hit in the innermost layer of the pixel detector are
classified as {\em converted} photon candidates.
The final energy measurement, for both converted and unconverted
photons, is made using only the calorimeter, with a cluster size that
depends on the photon classification. In the barrel, a cluster
corresponding to 3$\times$5 ($\eta\times\phi$) cells in the second
layer is used for unconverted photons, while a cluster of 3$\times$7
($\eta\times\phi$) cells is used for converted photon candidates to
compensate for the opening between the conversion products in the
$\phi$ direction due to the magnetic field. 
In the end-cap, where the cell size along $\theta$ is smaller than in the
barrel and the conversion tracks are closer in $\phi$ because of
the smaller inner radius of the calorimeter, a cluster size of 5$\times$5 is
used for all candidates.
A dedicated energy calibration~\cite{ATLAS_CSC} is then applied separately for
converted and unconverted photon candidates to account for upstream
energy loss and both lateral and longitudinal leakage.
Both unconverted and converted photon candidates are considered for
this measurement.
Photons reconstructed near regions of the calorimeter affected by readout or
high-voltage failures are not considered, eliminating around 5\% of the
selected candidates.
Events with at least one photon candidate with transverse energy
$\ET^\gamma>25$ GeV
and pseudorapidity $|\eta^\gamma|< 1.37$ are selected.
Photons are selected using the same shower-shape and 
isolation variables discussed in Refs.~\cite{promptphotonpaper_2010} 
and~\cite{photonPUB_2011}. 
The selection criteria on the shower-shape variables are independent of the
photon candidate's transverse energy, but vary as a function of the
photon reconstructed pseudorapidity, to take into account variations
in the total thickness of the upstream material and in the calorimeter
geometry. They are optimized independently for unconverted and
converted photons to account for the different developments of the
showers in each case. Applying these selection criteria
suppresses backgrounds from jets misidentified as photons.
The photon transverse isolation energy \Etiso\ is required to be
lower than 3 GeV. 
Less than $0.2\%$ of events have more than one
photon candidate passing the selection criteria. 
In such events the leading-\ET~photon is retained.

\subsection{Jet selection}
\label{subsec:jet_selection}
Jets are reconstructed starting from three-dimensional topological
clusters built from calorimeter cells,
using the infrared- and collinear-safe
anti-$k_t$ algorithm~\cite{antikt}
with a radius parameter $R = 0.4$.
The jet four-momenta are constructed from a sum over their
constituents, treating each as an $(E,\vec{p})$ four-vector with 
zero mass.
The jet four-momenta are then recalibrated using a jet energy scale
correction as described 
in Ref.~\cite{jes_uncertainty2}.
The calibration procedure corrects for instrumental effects, 
such as inactive material and noncompensation, as well as for
the additional energy due to multiple $pp$ interactions 
within the same bunch crossing ({\em pile-up}).
Jets with calibrated transverse momenta greater than 20 GeV are 
retained for this measurement.

To reject jets reconstructed from calorimeter signals not originating
from a $pp$ collision, the same jet quality criteria used in
Ref.~\cite{jes_uncertainty2}
are applied here.
These cuts suppress fake jets from calorimeter noise, cosmic rays and
beam-related backgrounds.

Jets overlapping with the candidate photon, or with an isolated 
electron produced from $W$ or $Z$ decay, are not considered. For this reason, 
if the jet axis is within a cone of radius 0.3 around the photon, the jet is 
discarded. Similarly, if the jet axis is within a cone of radius 0.3 around
any electron that passes the tight identification 
criteria~\cite{electronperf_2011} and that has
calorimeter isolation, $\Etiso$, less than 4 GeV, the jet is discarded.

The average jet multiplicity after the previous requirements is between
1.3 and 2.0, increasing with $\ET^\gamma$. 
In events with multiple jet candidates, the leading-$\pT$~jet is chosen.
In order to retain the event, the leading jet is required to have
rapidity $|y^{\rm jet}|<4.4$.
The leading jet axis is also required not to lie within a cone of radius $R
= 1.0$ around the photon direction.

The contamination in the selected sample from pile-up jets is estimated
to be negligible, which is consistent with the low pile-up conditions of
the 2010 data-taking, when, on average, only two minimum-bias events 
per bunch crossing are expected. 

\subsection{Distribution of photon transverse energy in selected events}
\begin{figure}[!htb]
  \centering
  \includegraphics[width=1.0\columnwidth]{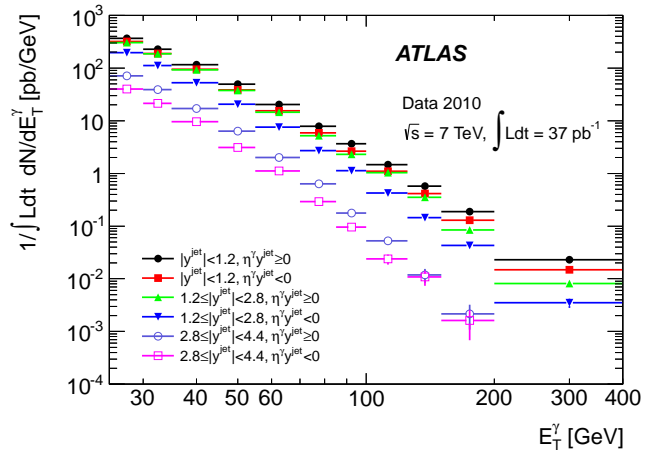}
  \caption{Transverse energy distribution of photon candidates
    in photon + jet events selected in the 2010 ATLAS data, before
    background subtraction.
    The distribution is normalized by the integrated luminosity and the
    transverse energy bin width. Events with $\ET^\gamma\leq 45$ GeV have
    been collected with the (prescaled) 20 GeV photon trigger.
    Events with $\ET^\gamma>45$ GeV have been collected with the
    (unprescaled) 40 GeV photon trigger.} 
  \label{fig:et_distribution}
\end{figure}

The number of events after photon and jet selections is 213\,003. 96\,314 events have been
collected with the 20 GeV trigger and have
$25$~GeV$<\ET^\gamma\leq 45$~GeV, 116\,689 events 
have been collected with the 40 GeV trigger and have $\ET^\gamma>45$~GeV.
In 57\% of the events the jet is central (32\%/25\% are in the same/opposite-sign configuration), 
in 37\% of the events the jet is forward (24\%/13\% are in the same/opposite-sign configuration), and 
in 6\% of the events the jet is very forward (4\%/2\% are in the same/opposite-sign photon). 
The photon candidate is reconstructed as unconverted in 68\% of the
events and as converted in the remaining 32\%.
The transverse energy distribution of the photon candidates 
in the selected sample is shown in Fig.~\ref{fig:et_distribution}.

\section{Background Subtraction and Signal Yield Estimation}
\label{sec:Purity}
A non-negligible residual contribution of background is
expected in the selected photon + jets sample, even after the application of
the tight identification and isolation requirements.
The dominant background 
is composed of dijet events in which one jet is misidentified as a
prompt photon, with a tiny contribution from diphoton and $W/Z$+jets
events.
In more than 95\% of background dijet events, the misidentified jet contains
a light neutral meson that carries most of the jet energy and decays
to a collimated photon pair.
The background yield in the selected sample is
estimated {\em in situ} using a two-dimensional sideband technique as in
Ref.~\cite{promptphotonpaper_2010}
and then subtracted from the observed yield.
In the background estimate, the photon is classified as:
\begin{itemize}
\item Isolated, if $\Etiso<3$ GeV;
\item Nonisolated, if $\Etiso>5$ GeV;
\item Tight, if it passes the tight photon identification
criteria;
\item Nontight, if it fails at least one of the
tight requirements on four shower-shape variables computed from the energy
deposits in a few cells of the first layer of the electromagnetic
calorimeter, but passes all the other tight identification criteria.
\end{itemize}
In the two-dimensional plane~\cite{promptphotonpaper_2010} formed 
by the photon transverse isolation energy and the photon tight 
identification variable, we define four regions: 
\begin{itemize}
\item{$A$}: the {\em signal} region, containing tight, isolated photon
candidates.
\item{$B$}: the {\em nonisolated} background control region, containing
tight, nonisolated photon candidates.
\item{$C$}: the {\em nonidentified} background control region,
containing isolated, nontight photon candidates.
\item{$D$}: the background control region containing nonisolated,
nontight photon candidates.
\end{itemize}

The signal yield $N_A^{\rm sig}$ in region $A$ is estimated from 
the number of events in the four regions, $N_K$ ($K \in
\{A,B,C,D\}$), through the relation 
\begin{equation}
\label{eq:sigyield_leakage}
   N_A^{\rm sig} = N_A - (N_B-c_BN_A^{\rm sig})\frac{(N_C-c_CN_A^{\rm sig})}{(N_D-c_DN_A^{\rm sig})}\ ,
\end{equation}
where $c_K\equiv {N^{\rm sig}_K}/{N^{\rm sig}_A}$
 are {\em signal leakage fractions} that can be extracted
from simulated signal event samples. Equation~\ref{eq:sigyield_leakage} leads to a
second-order polynomial equation in $N_A^{\rm sig}$ that has only one 
physical ($N_A^{\rm sig}>0$) solution.
The only hypothesis underlying Eq.~\ref{eq:sigyield_leakage} is that 
the isolation and identification variables are uncorrelated in
background events. 
This assumption has been verified both in background simulated samples,
and in data in the background-dominated region of $\Etiso > 7$ GeV.
This method was found to return signal yields consistent with the
generated ones using a cross section weighted combination of simulated
signal and background samples.

\begin{figure*}[!htbp]
  \centering
  \includegraphics[width=1.0\columnwidth]{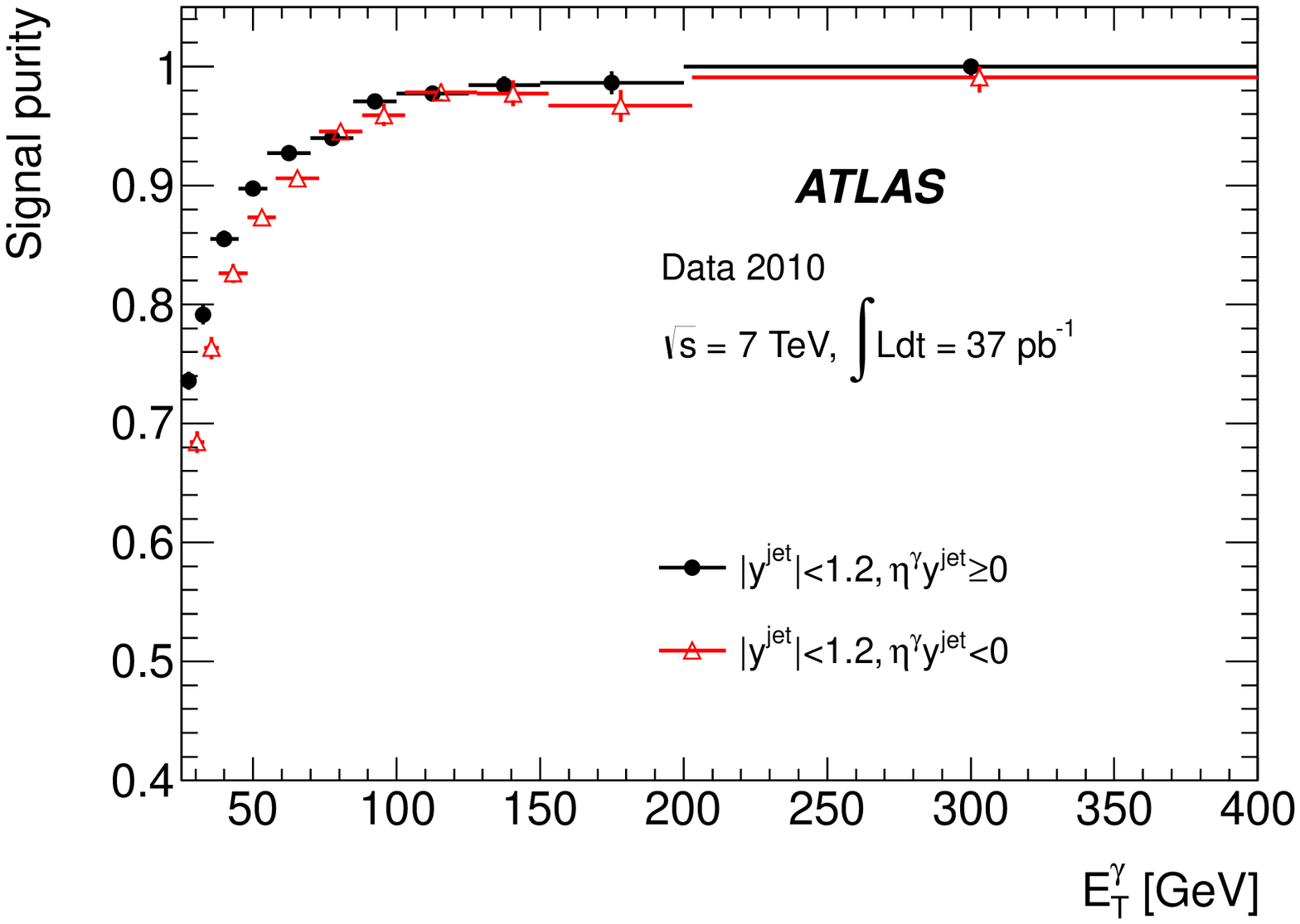}
  \includegraphics[width=1.0\columnwidth]{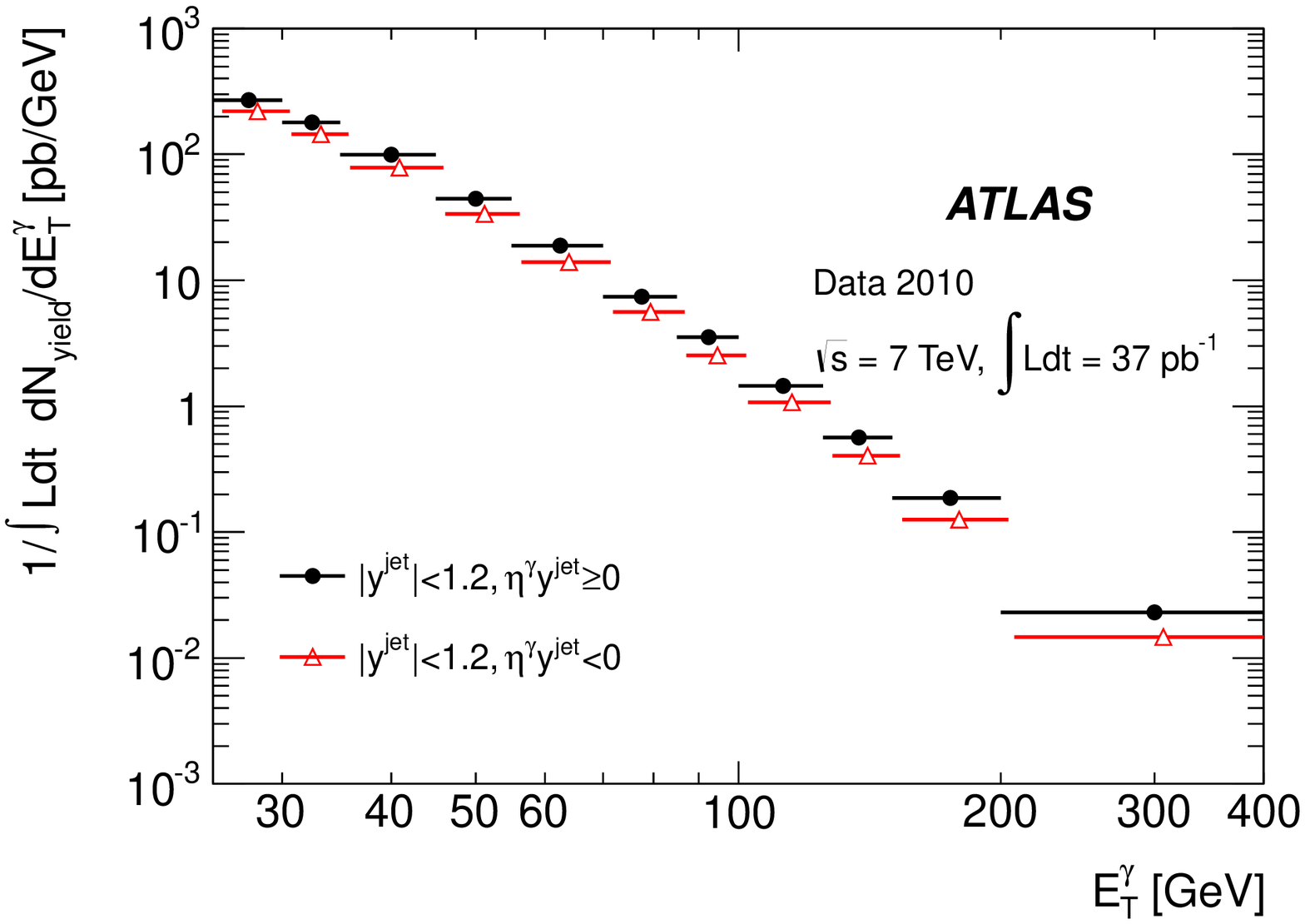}
  \includegraphics[width=1.0\columnwidth]{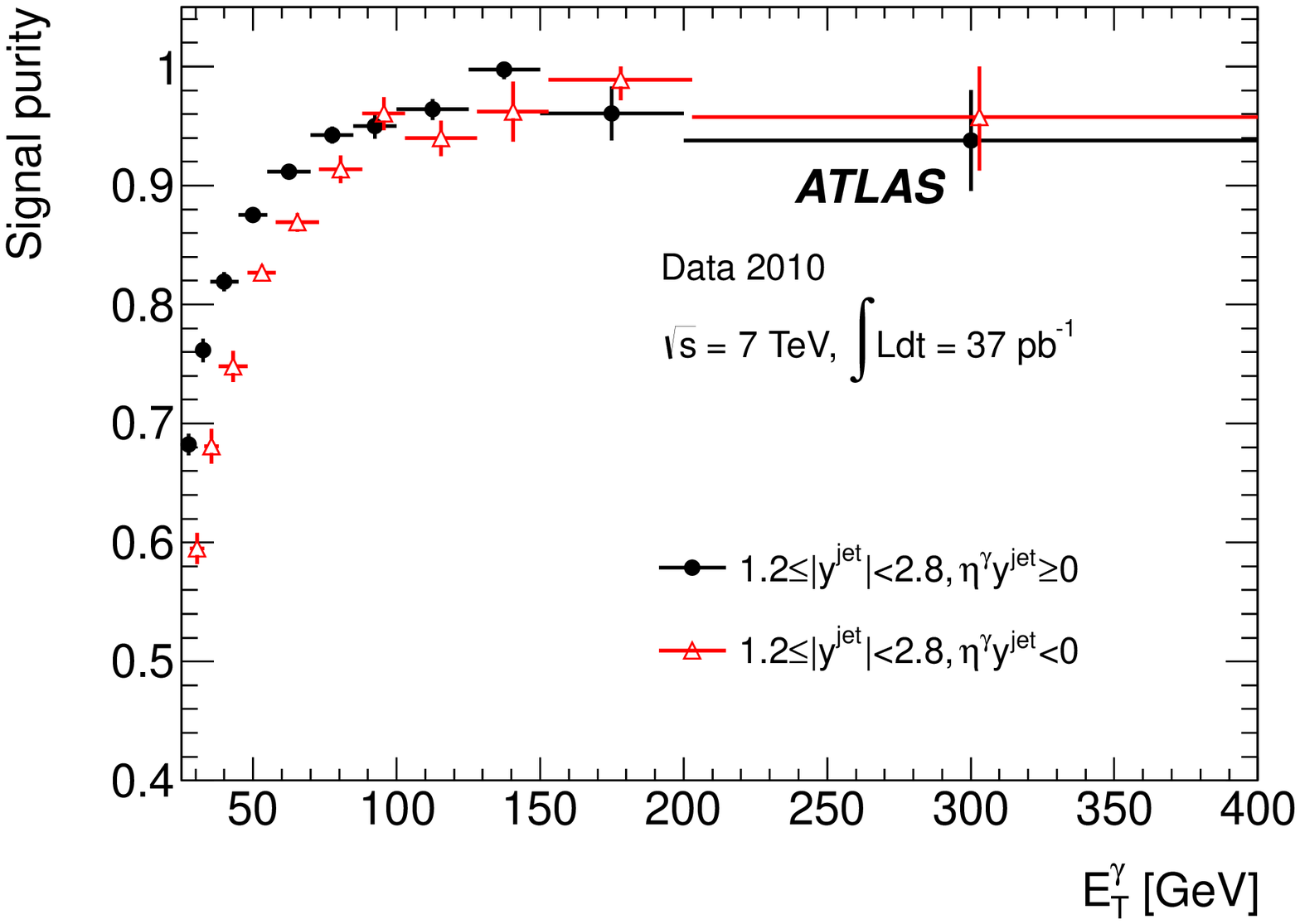}
  \includegraphics[width=1.0\columnwidth]{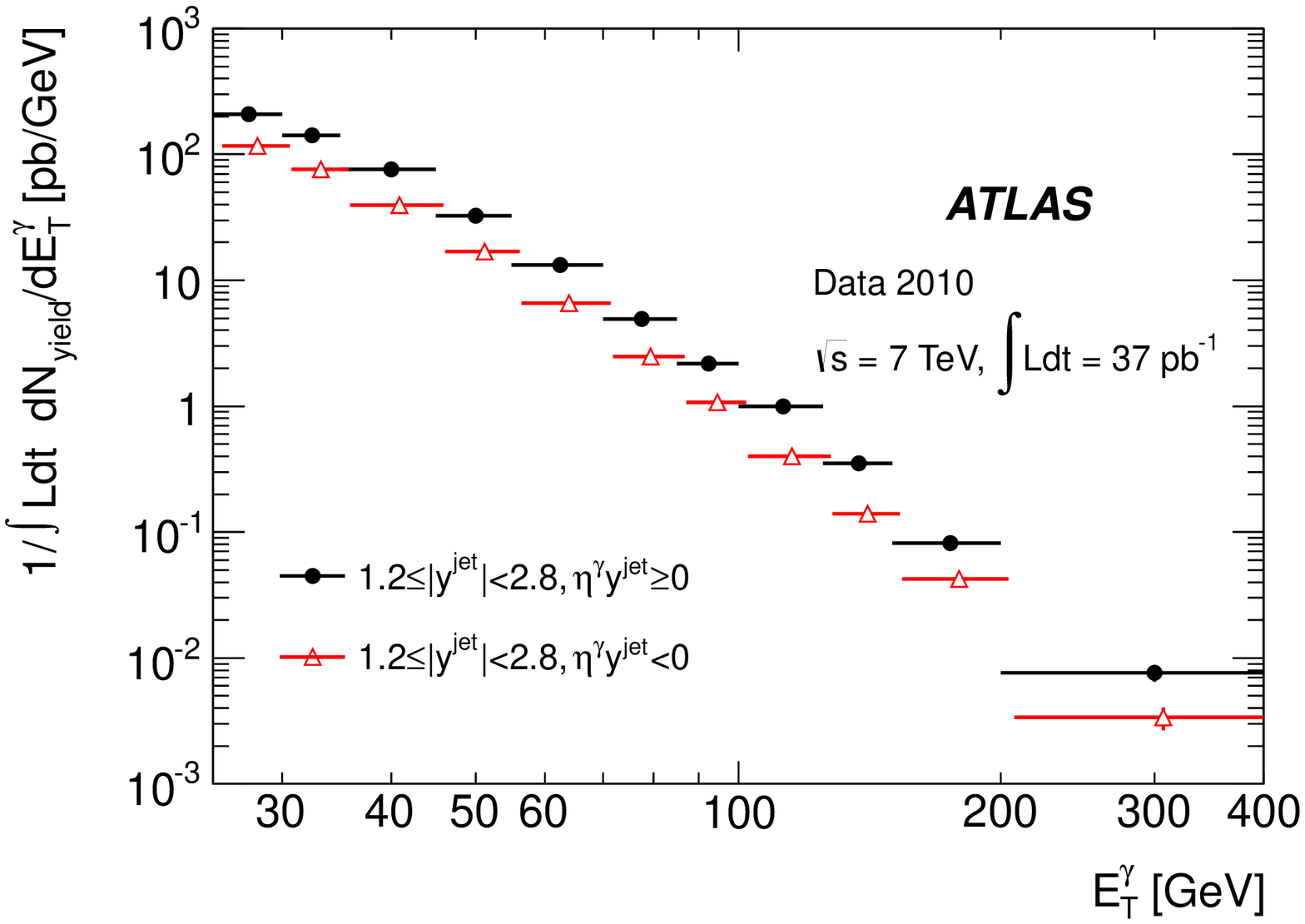}
  \includegraphics[width=1.0\columnwidth]{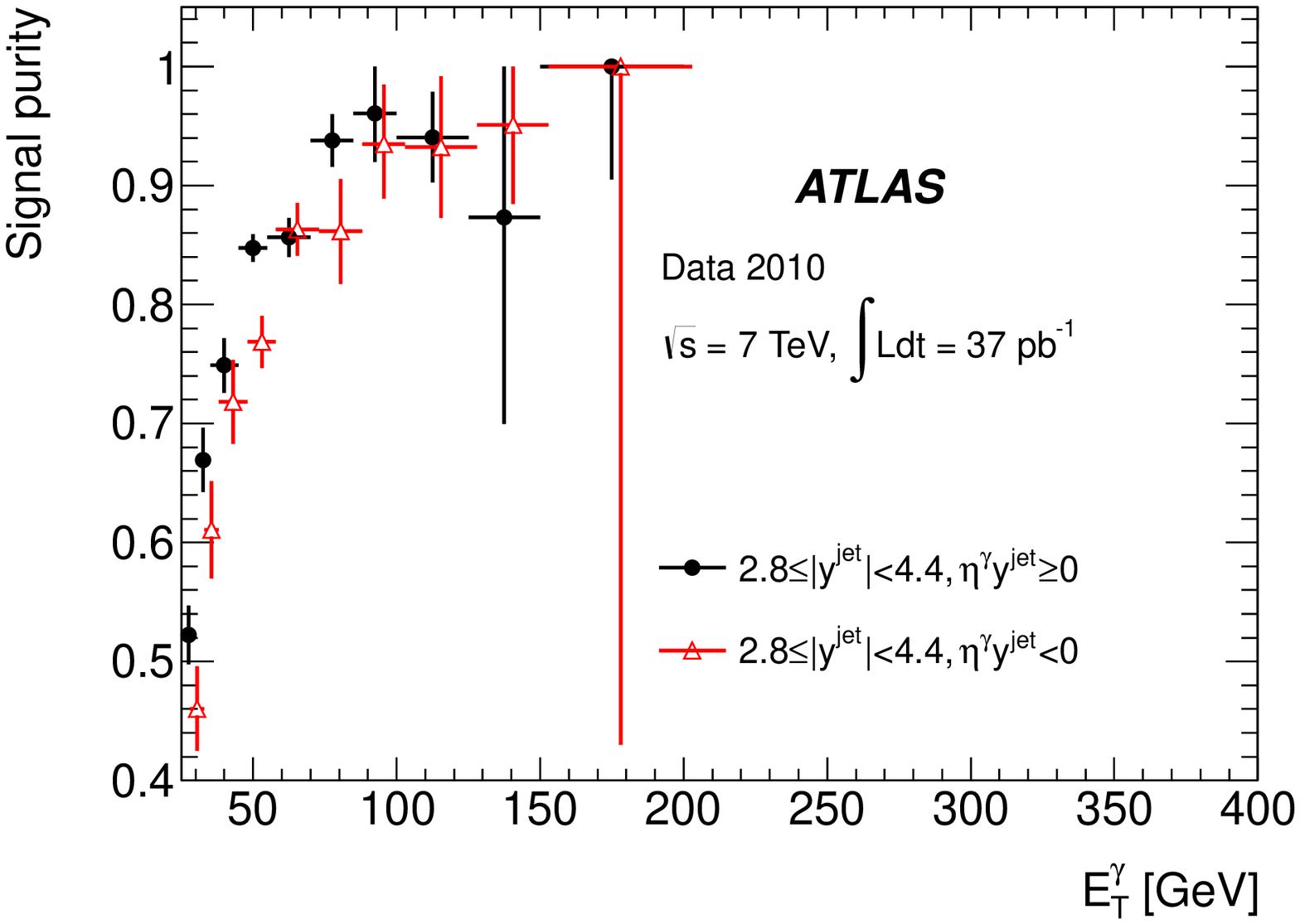}
  \includegraphics[width=1.0\columnwidth]{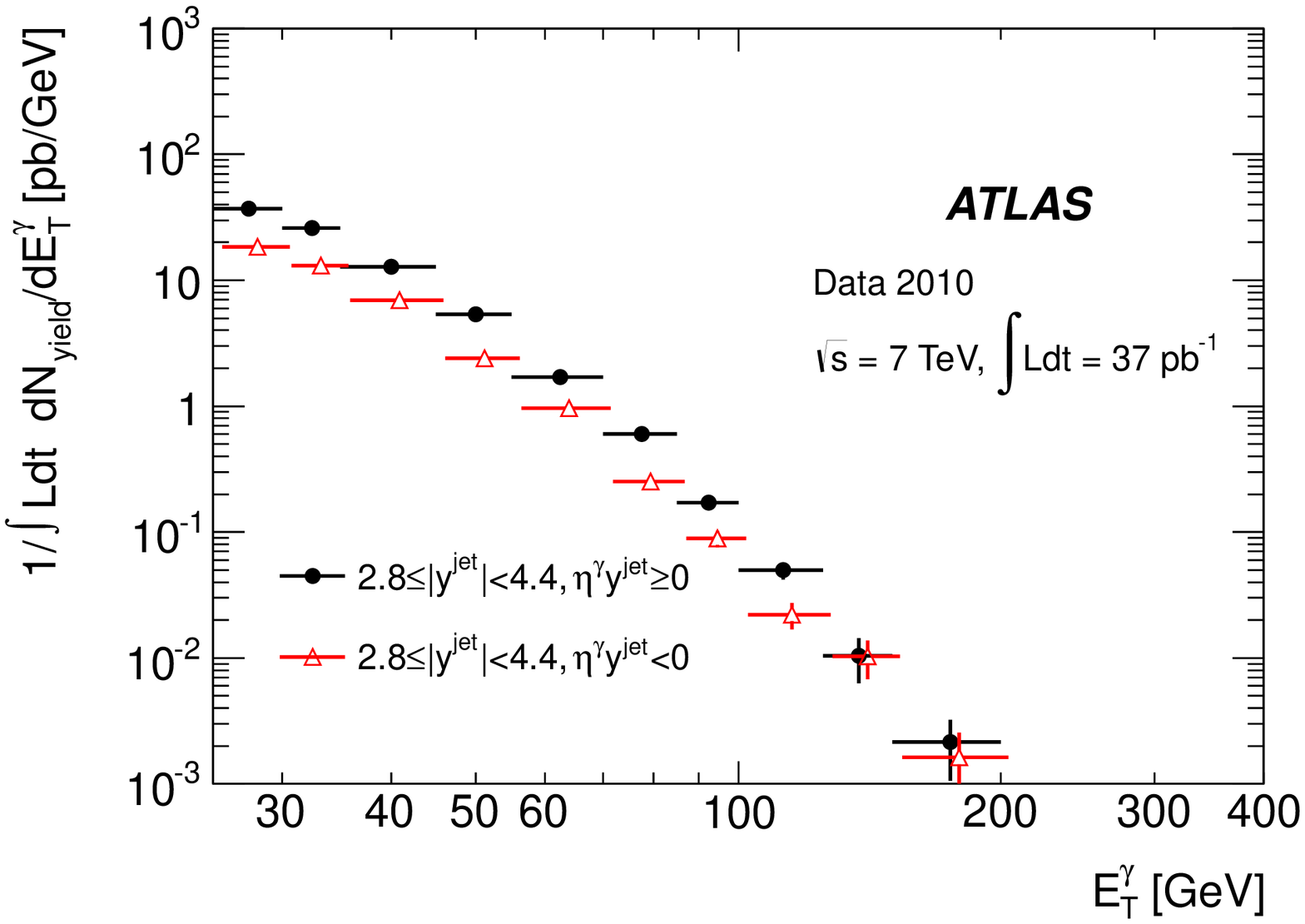}
  \caption{Estimated signal purity (left column) and signal yield normalized 
  by bin width and integrated luminosity (right column) in data as a 
  function of the photon transverse energy, for the same-sign angular 
  configurations (full circles) 
  and the opposite-sign angular configurations (open triangles).
  A small horizontal displacement has been added to the points
  corresponding to the opposite-sign configurations, so that the error
  bars are clearly shown. The errors are statistical only. 
  Top row: central jet. Middle row: forward jet. Bottom row: very forward jet.}
  \label{fig:purity_and_yield_vs_et}
\end{figure*}

The resulting signal purity and signal yield
as a function of the
photon candidate transverse energy for the six photon and jet angular
configurations are shown in Fig.~\ref{fig:purity_and_yield_vs_et}.
The signal purity typically increases from between 50\% and 70\% at
$\ET^\gamma=25$~GeV to above 95\% for $\ET^\gamma>150$~GeV.
The effect of the non-negligible signal leakage in the background
control regions ($c_K\ne 0$) increases the measured purity by
5-6\% at $\ET^\gamma=25$~GeV and $\approx 2$\% at
$\ET^\gamma>150$~GeV compared to the purity estimated
assuming negligible signal in the background regions.

\section{Signal Efficiency and Cross Section Measurement}
\label{sec:Efficiency}
The combined signal trigger, reconstruction, and selection efficiency is
evaluated from the simulated signal samples described in
Sec.~\ref{subsec:simulated_data}, which include leading order
$\gamma$ + jet events from both 
hard-scattering (hard subprocesses $q g \rightarrow q \gamma$ and $q
\bar{q} \rightarrow g \gamma$)
and from quark bremsstrahlung in QCD dijet events. 
For each of the six angular configurations, efficiency matrices ($\Lambda_{ij}$)
are constructed, with the indices $i$ and $j$ corresponding to 
reconstructed and true photon transverse energy intervals, respectively.
The efficiency matrices account both for
trigger, reconstruction, photon identification efficiencies and for
migrations between different bins of the true and reconstructed photon
transverse energies due to resolution effects.
The matrix elements are determined from the ratios of two quantities. 
The denominators are defined in the following way:

\begin{itemize}

\item{} The leading truth-level signal photon
within the acceptance ($|\eta^\gamma_{\rm true}| \leq 1.37$) is selected.

\item{} Truth jets are reconstructed using the anti-$k_t$
algorithm with a radius parameter $R=0.4$ on all the particles
with proper lifetime longer than 10~ps, including
photons, and the leading truth jet is selected among those with axis
separated from the photon direction by $\Delta R>0.3$.
The leading photon and the leading jet are required to be separated by
$\Delta R >$ 1.0. 

\item{} To retain the event the true leading photon is required to have
$E^\gamma_{\rm T,true} > 20$~GeV and
to have a truth-particle-level isolation (computed from the true four-momenta
of the generated particles inside a cone of radius 0.4 around the
photon direction) $E^{\rm iso}_{\rm T,true}<4 \GeV$. This
truth-particle-level cut has been determined on PYTHIA photon + jet samples 
to match the efficiency of the experimental isolation cut
at 3 GeV (more details can be found in Ref.~\cite{promptphotonpaper_2010}).
In this case, the same underlying event subtraction procedure used on data
has been applied at the truth level.
In addition, the leading truth jet is required to have $E^{\rm jet}_{\rm
T,true} > 20$~GeV and $|y^{\rm jet}_{\rm true}| < 4.4$.
At the truth level the minimum $E^\gamma_{\rm T,true}$ is set to 20~GeV
to account for possible migrations of photons with true
transverse energy below 25~GeV in the reconstructed transverse 
energy intervals above 25~GeV.

\end{itemize}
The numerators are determined by applying the selection criteria
described in Sec.~\ref{sec:Selection} to the simulated signal
samples. 
Since the simulation does not describe accurately the electromagnetic
shower profiles, a correction factor for each simulated shape variable is
applied to better match the data.
We require the reconstructed isolation energy to be less than 3
$\GeV$. 
As for the truth level, photons are allowed to have a $E^\gamma_{\rm
T,reco}>20$ GeV. 
The reconstructed photon is required to match the truth photon within a
cone of radius 0.4 while the reconstructed jet is required to match the
truth jet in a cone of radius 0.3.
Events which pass the selection at the reconstruction level but fail
it at the truth level are 
properly accounted for in the normalization.

The 
event selection efficiency typically rises from 
50\% to 80\% as a function of $\ET^\gamma$. 
An inefficiency of around 15\% is due to the acceptance loss 
originating from a few inoperative optical links in the calorimeter readout 
and from the isolation requirement. 
An inefficiency decreasing from 20-25\% for $\ET^\gamma = 25$~GeV 
to almost zero at high $\ET^\gamma$ originates from the shower-shape
photon identification selection. 

The differential cross section as a function of the photon true
transverse energy $\ET^{\rm true}$ is computed in each bin \textit{i} of
$\ET^{\rm true}$ and for each angular configuration \textit{k} as:
\begin{equation}
\frac{d\sigma^{k}_{i}}{d\ET^{\rm true}}=\frac{N^{{\gamma},{\rm
true,isol},k}_{i}}{\int Ldt~\Delta E^{\rm true}_{{\rm T},i}}\ ,
\end{equation}
where $N^{{\gamma},{\rm true,isol},k}_{i}$ is
the number of events containing a true isolated photon and
hadronic jets, 
in which the true photon transverse energy is in 
bin \textit{i} and the angular configuration formed 
by the leading photon and jet is \textit{k}.
This number is related to the observed number of
events passing the analysis cuts through the efficiency matrices $\Lambda_{ij}$:

\begin{equation}
  \label{eq:eff_unf}
  N^{{\gamma},{\rm reco,isol},k}_{i} = \sum_{j} \Lambda_{ij}
  N^{{\gamma},{\rm true,isol},k}_{j}
\end{equation}

The unfolding procedure allows the reconstruction of the true number of 
events from the measured distribution, taking into account the
measurement uncertainties due to statistical fluctuations in the
finite measured sample. 
The simplest unfolding method is the basic
\textit{bin-by-bin} unfolding, which corrects the
observed cross section in bin \textit{i} with the efficiency obtained from
the ratio of 
selected events to truth events having the photon
with reconstructed and true $\ET$ in bin \textit{i}.
A more sophisticated
method which properly accounts for migrations between bins is based on the
repeated (iterative)
application of Bayes's theorem~\cite{bayes_unfolding}.
The differences in the measured cross section for the
two methods are a few percent for events with a central
or forward jet and slightly higher for events with a very forward jet.
Since the differences are within the statistical errors of the methods, we
used the bin-by-bin method for these results.

\section{Systematic Uncertainties}
\label{sec:xsection}
We have considered the following sources of systematic uncertainties
in the cross section measurement (see Appendix~\ref{app:systematics} 
for tables detailing the uncertainties in each 
$\ET^{\gamma}$ bin and each angular configuration): 

\begin{itemize}

\item Simulation of the detector geometry.
The presence of material in front of the calorimeter
affects the photon conversion rate
and the development of electromagnetic showers. Therefore the cross
section measurement uncertainty depends on the accuracy of the detector
simulation. The nominal simulation may underestimate
the actual amount of material in front of the calorimeters.
To quantify the effect of more material on the cross section,
the full analysis is 
repeated using a detector simulation with a conservative
estimate of additional material in front of the 
calorimeter~\cite{electronperf_2011}. In this
case the photon identification and reconstruction efficiencies are lower
than in the nominal case. The increase in cross section is assigned as 
a positive systematic uncertainty.
In the central and forward jet configurations the systematic
uncertainty
varies from 5\% to 8\% for photons with 25~GeV$<\ET^\gamma\leq 45$~GeV
and from 1\% to 5\% for $\ET^\gamma > 45$~GeV. In the very forward jet
configurations the uncertainty is similarly estimated to range
from 10\% to 23\%.

\item Photon simulation. In order to take into account
  the uncertainty on the event generation and the parton shower model, four
  additional samples are used: {\tt PYTHIA} or {\tt HERWIG} samples
  containing only hard-scattering photons and {\tt PYTHIA} or {\tt
    HERWIG} samples containing only photons from quark
  bremsstrahlung. The analysis is repeated using these samples, and 
  the largest positive and negative deviations from
  the nominal cross section are taken as systematic uncertainties.
  The deviations are mainly positive,
  varying from 4\% to 16\% depending on
  $\ET^{\gamma}$ or the angular configuration. 

\item Jet and photon energy scale and resolution
  uncertainties. The cross section uncertainty is determined 
  by varying the electromagnetic and
  the jet energy scales and resolutions within their 
  uncertainties~\cite{jes_uncertainty2,electronperf_2011}.
  The effect on the cross section is found to be negligible, with the 
  exception of the effect of the jet energy scale uncertainty, which
  affects mainly the first $\ET^{\gamma}$ bin due to the  
  efficiency of the 20~GeV threshold on $\pT^{\rm jet}$. For the angular
  configurations including one central or one forward jet this effect
  is 3\% to 7\%, for the configurations containing one very forward jet it is
  9\% to 20\%.

\item Uncertainty on the background correlation in the two-dimensional
  sidebands method. 
  The isolation and identification variables are assumed 
  to be independent for fake photon candidates. This
  assumption was verified using both data and simulated background
  samples and was found to be valid within a 10\% uncertainty for
  configurations including a central or a forward jet and within a 25\%
  uncertainty for configurations including a very forward jet. 
  The cross section is recomputed accounting for these possible
  correlations in the background
  subtraction~\cite{promptphotonpaper_2010},
  and the difference with the nominal result is taken as a systematic
  uncertainty.
  This procedure gives a systematic uncertainty on the cross section of 3\%
  and 6\% in the first $\ET^{\gamma}$ bin for these groups of
  configurations respectively. This uncertainty decreases rapidly with 
  increasing $\ET^{\gamma}$, being proportional to $1-P$, 
  where $P$ is the signal purity.

\item Background control regions definition in the 
  two-dimensional sidebands method.
  The measurement is repeated using a different set of
  background identification or isolation criteria in the purity
  calculation, and the difference between the new cross section and the
  nominal result is taken as a systematic uncertainty.
  For background identification, three or five shower-shape variables
  are reversed instead of four as in the nominal case (more
  details can be found in Ref.~\cite{promptphotonpaper_2010}). 
  The deviations on the cross section range from 5\% in the central jet
  configurations to 12\% in the forward jet configurations, all decreasing
  with increasing $\ET^{\gamma}$. 
  Varying the isolation cut by $\pm 1$~GeV results in 
  less than 1\% difference in the cross section.
  
\item Data-driven correction to the photon efficiency.
  The simulated photon shapes in the calorimeter have been corrected
  in order to improve the agreement with the data. 
  The systematic uncertainty related to the correction procedure is
  computed using different simulated photon samples and a different
  simulation of the ATLAS detector and 
  is estimated to be of the order of 1\% to 4\% in the first
  $\ET^{\gamma}$ bin and lower than 1\% elsewhere~\cite{promptphoton_confnote_2011}.
  
\item Uncertainty on the trigger efficiency. The trigger efficiency in the
  simulation is consistent with the one measured in data, using a
  bootstrap method, within the
  total uncertainty of the {\em in situ} measurement (0.6\%
  uncertainty for
  $\ET^{\gamma}\leq 45$~GeV and 0.4\% for $\ET^{\gamma}>45$~GeV). 
  These uncertainties are added to the total systematic uncertainty 
  on the cross section.

\item Uncertainty on the jet reconstruction efficiency. The simulation
  is found to reproduce data jet reconstruction efficiencies to better
  than 2\%~\cite{jetconfnote_2011}. A 2\% systematic uncertainty to the
  cross section is assigned.

\item Uncertainty on the simulated jet multiplicity. The LO generators
  used to estimate the signal efficiencies do not reproduce precisely
  the jet
  multiplicity observed in data, and the signal efficiency could
  depend on the multiplicity. Reweighting the simulation in
  order to reproduce the jet multiplicity observed in data changes the
  cross section by less than 1\%, which is taken as a systematic
  uncertainty.

\item Uncertainty on the integrated luminosity. It has been determined
  to be 3.4\%~\cite{lumipaper,ATLAS-CONF-2011-011}.

\item Isolated electron background. Possible backgrounds may arise
  from $W$+jets where the $W$ decays into an
  electron misidentified as photon, and $W+\gamma$ where the $W$ 
  decays into an electron misidentified as a jet. Additional backgrounds
  may originate from $Z\rightarrow ee$
  where an electron may be misidentified as a photon, and combined
  with the jet arising from the misidentification of the other electron 
  or with a jet from the rest of the event (in $Z$+jets).
  Using simulated samples of these processes, scaled to their
  cross sections measured 
  in~\cite{ATLAS_Wjet, ATLAS_Wgamma, ATLAS_WZ}, the total isolated
  electron background is 
  estimated to be less than 1.5\% of the signal
  yield measured in data in each photon $\ET^{\gamma}$ bin. 
  Therefore an asymmetric systematic uncertainty 
  ($^{+0.0}_{-1.5}$)\% on the measured cross section is assigned.
\end{itemize}

The sources of systematic uncertainty discussed above are considered
as uncorrelated and thus the total systematic uncertainty (listed in
the tables in Appendix~\ref{app:xsmeasured}) is estimated by
summing in quadrature all the contributions.

\section{Theoretical Predictions}
\label{sec:Theory}
The expected production cross section of an isolated photon in 
association with jets as a function of the photon transverse energy
$\ET^\gamma$ is estimated 
using {\tt JETPHOX} 1.3~\cite{jetphox1}.
{\tt JETPHOX} is a parton-level Monte Carlo generator which implements
a full NLO QCD calculation of both the direct and fragmentation
contributions to the cross section.
A parton-level isolation cut, requiring a total transverse energy
below 4 GeV from the partons produced with the photon inside a cone of
radius $\Delta R=0.4$ in $\eta\times\phi$
around the photon direction, is used for this computation. 
The NLO photon fragmentation function~\cite{Bourhis:2000gs} and the
{\tt CT10} parton density functions~\cite{cteq} are used. 
The nominal renormalization ($\mu_R$), factorization ($\mu_F$) and
fragmentation ($\mu_f$) scales are set to the photon transverse
energy $\ET^\gamma$. Jets of partons are reconstructed by using an
anti-$k_{T}$ algorithm with a radius parameter $R = 0.4$. 
The same transverse momentum and rapidity criteria applied in the
measurement to the reconstructed objects are used in the {\tt JETPHOX}
generation for the photon and the leading-$\pT$ jet. As for data, the
event is kept if the two objects are separated by $\Delta R>1.0$
in $\{\eta,\phi\}$. With this setup the fragmentation
contribution to the total cross section decreases as a function of
$\ET^\gamma$, from 10\% to 1.5\% for the same-sign, central jet
configuration while it varies from 22\% to 2.5\% in the same-sign, very
forward jet configuration. In the opposite-sign configurations the
fragmentation contribution
is 20\% to 50\% (depending on $\ET^\gamma$ and the
jet rapidity) higher than in the corresponding
same-sign configurations.

The {\tt JETPHOX} cross section does not include underlying event,
pile-up or hadronization effects.
While the ambient-energy density correction of the photon isolation
removes the effects from underlying event and pile-up on the photon side,
potential differences between the {\tt JETPHOX} theoretical
cross section and the
measured one may arise from the application of the jet selection, in
particular the transverse momentum threshold of 20 GeV.
This cut is applied at parton-level in {\tt JETPHOX} while it is applied to
particle jets in the measured cross section and in the fully simulated
{\tt PYTHIA} and {\tt HERWIG} samples.

One effect of hadronization is to spread energy outside of the jet
area, so the jet \pT~will tend to be lower than that of the
originating parton(s); on the other hand, the underlying event adds
extra particles to the jet candidate and results in the increase of
the jet \pT.
To estimate these effects we use the simulated signal {\tt PYTHIA}
samples to evaluate the ratios of truth-level cross sections with and
without
hadronization and underlying event, and subsequently we 
multiply each bin of the 
{\tt JETPHOX} cross sections by these ratios.
These correction factors are smaller than 1 (around 0.9-0.95) at low
$\ET^\gamma$, indicating that the impact of hadronization on the jet
$\pT$ is more important than the extra energy added from the
underlying event and pile-up.
The correction factors are consistent with one at high $\ET^\gamma$.
This finding is in agreement with the expectations, since the photon and the jet 
transverse momenta are correlated and for large
$\ET^\gamma$ the $\pT>20$~GeV cut becomes fully efficient
both at parton- and particle-level.

The systematic uncertainties on the QCD cross sections computed with
{\tt JETPHOX} are estimated in the following way:
\begin{itemize}
\item The scale uncertainty is evaluated by fixing any two scales
to the nominal value and varying the third between 0.5 and 2.0 times 
the nominal value.
In addition the effect of the coherent scale variations where all
three scales are varied together is also taken into account. 
The envelope of the values obtained with the different scale configurations
is taken as a systematic uncertainty.
This leads to a change of the predicted
cross section between 15\% at low $\ET^\gamma$ and 10\% at high $\ET^\gamma$. 
\item The uncertainty on the cross section from the PDF uncertainty
has been obtained by varying the PDFs within the 68\% confidence level
intervals.
The corresponding uncertainty on the cross section varies
between 5\% and 2\% as $\ET^\gamma$ increases. 
Using a different set of PDFs, such as {\tt MSTW 2008}~\cite{mstw} 
or {\tt NNPDF 2.1}~\cite{nnpdf},
the computed cross sections vary always within the total systematic
uncertainty on the predicted cross section.
\item The uncertainty on the correspondence between parton-level and
particle-level isolation cut has been evaluated by varying the cut 
between 3 and 5~GeV. This variation
changes the predicted cross section by a few percent for the central
configuration but becomes more important for the forward and very forward
configurations, where the fragmentation contribution to the cross
section is larger.
\item The uncertainty on the hadronization and underlying event
corrections is estimated as the maximum spread of the correction
factors obtained from {\tt PYTHIA} using both the nominal and the Perugia
2010 tunes~\cite{PhysRevD.82.074018} and with {\tt HERWIG++} 2.5.1 with
the UE7000-2 tune~\cite{Bahr:2008pv}.
\end{itemize}

The expected cross sections with their full statistic and systematic 
uncertainties for all angular configurations under study are summarized
in Appendix~\ref{app:CStables_theory}.

\section{Comparison between data and theory}
\label{sec:DataTheoryComparison}
\begin{figure*}[!htbp]
  \centering
  \includegraphics[width=.8\columnwidth]{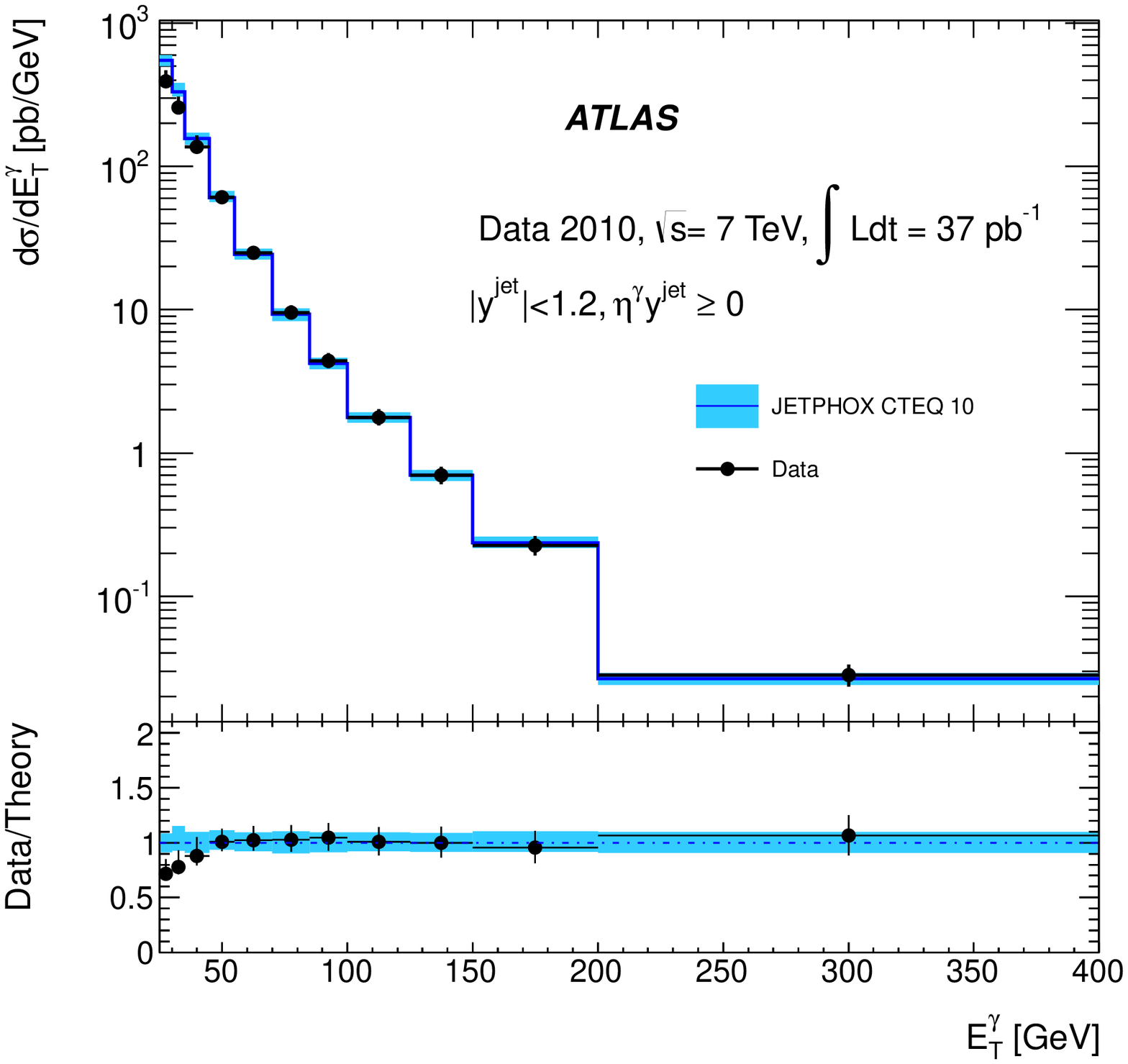}
\hspace{1cm}
  \includegraphics[width=.8\columnwidth]{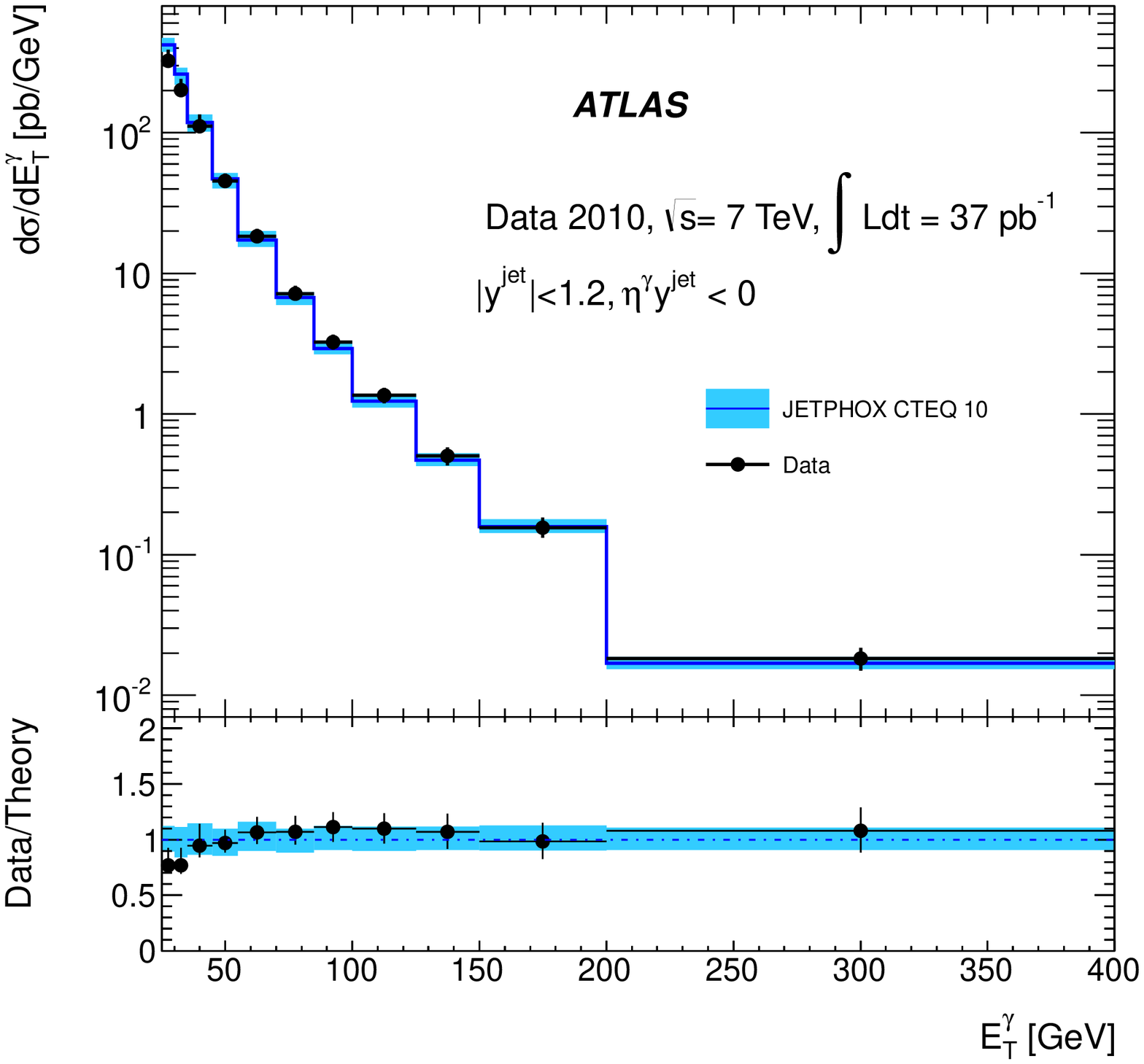}
  \includegraphics[width=.8\columnwidth]{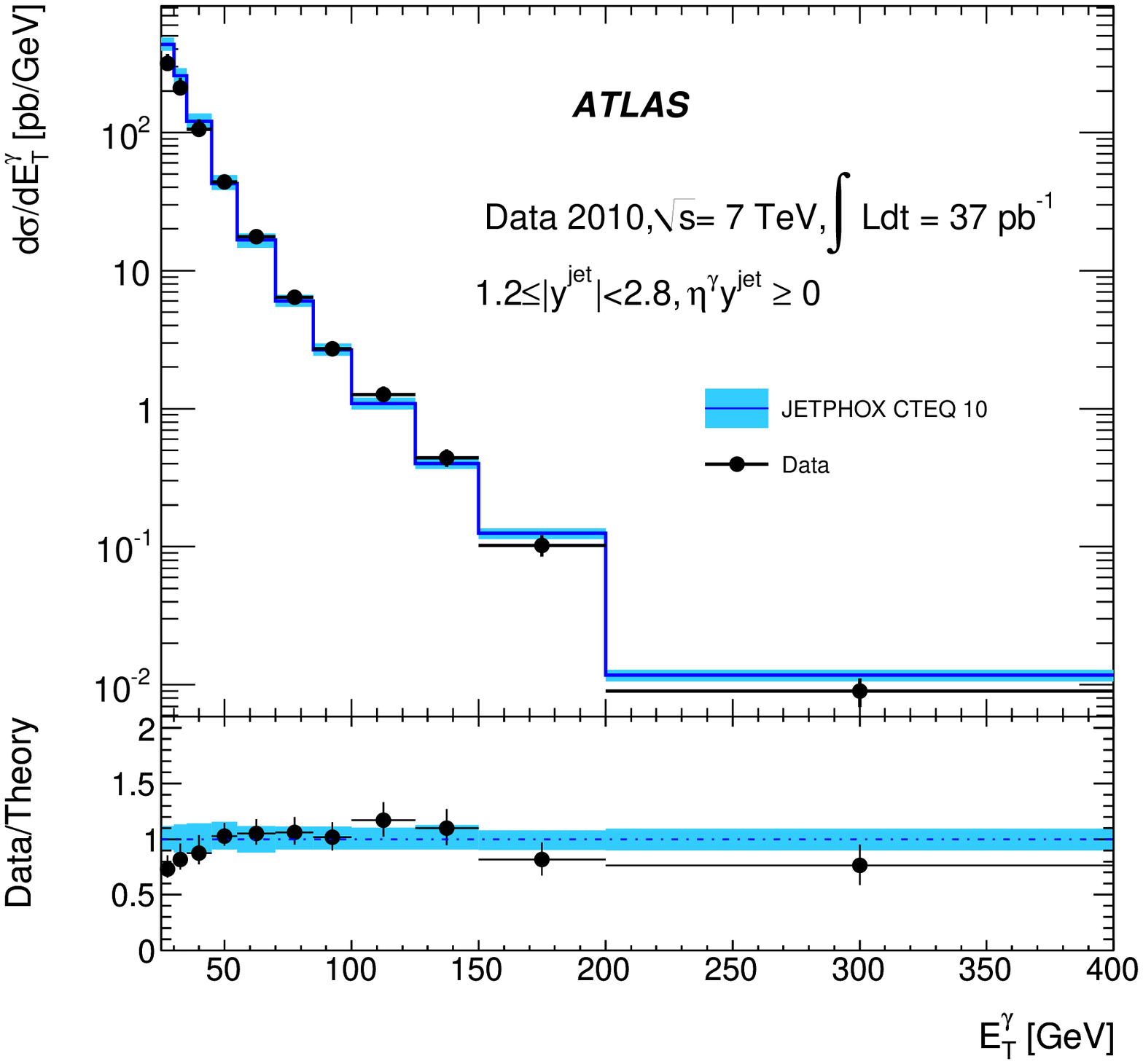}
\hspace{1cm}
  \includegraphics[width=.8\columnwidth]{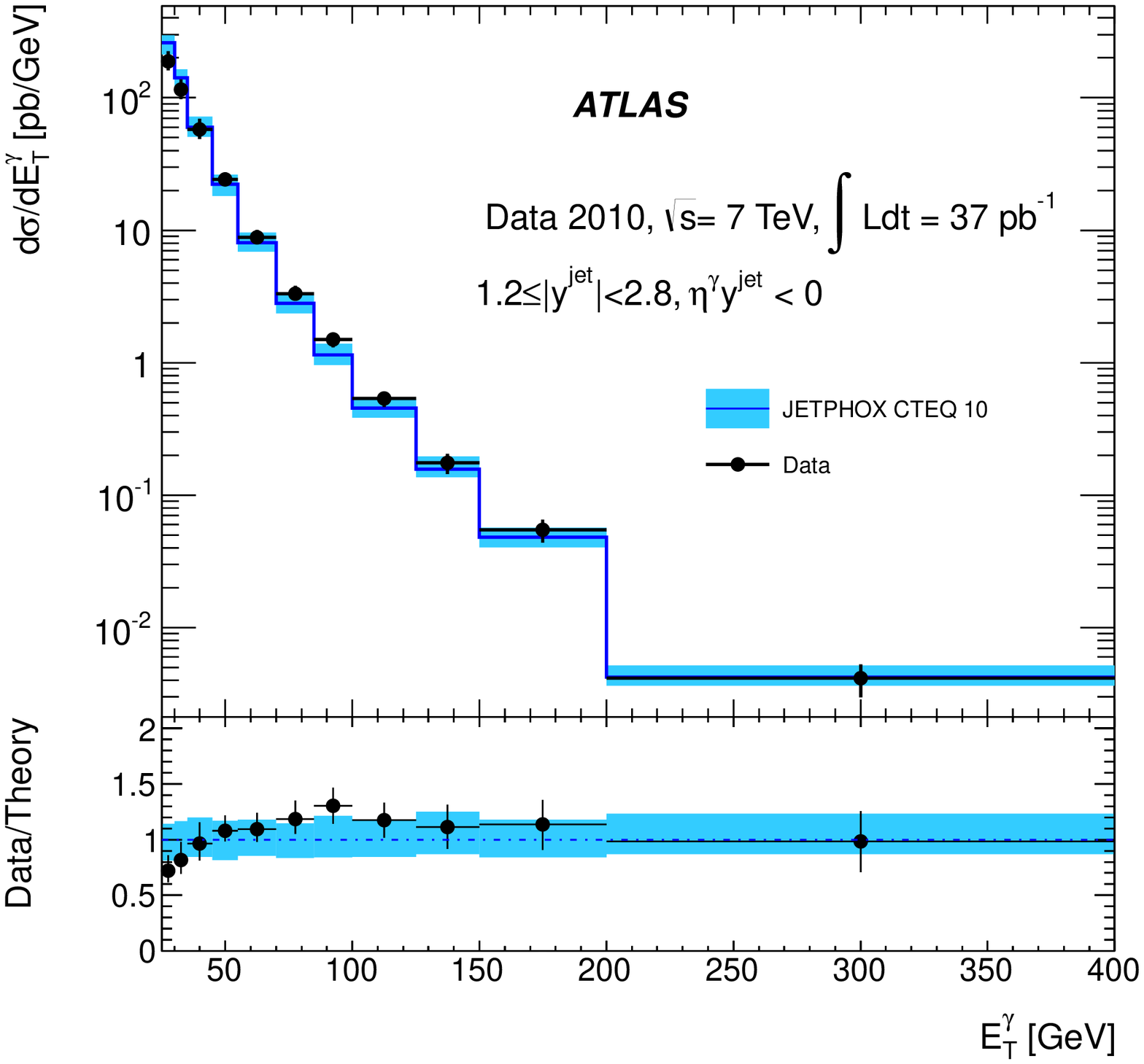}
  \includegraphics[width=.8\columnwidth]{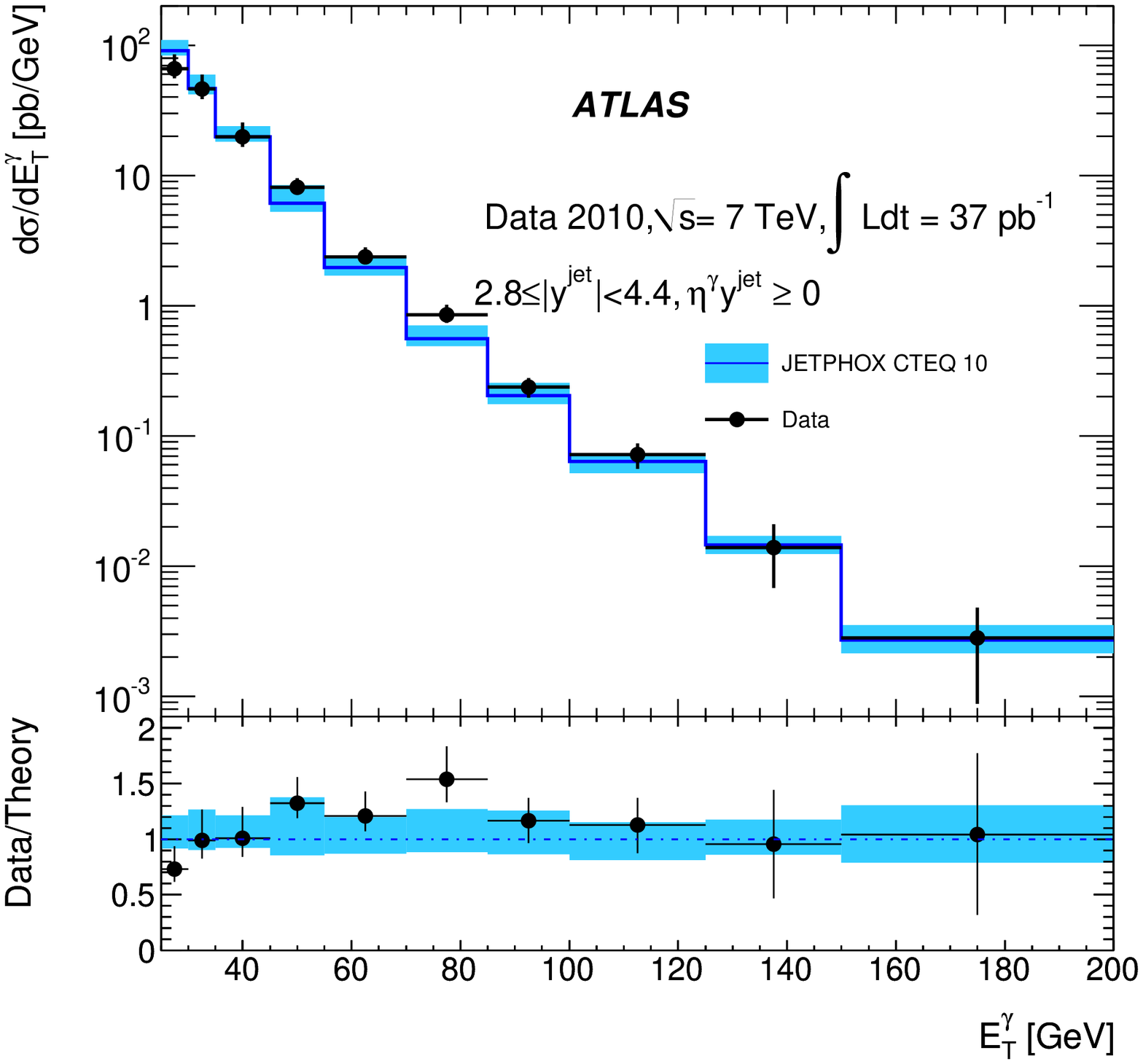}
\hspace{1cm}
  \includegraphics[width=.8\columnwidth]{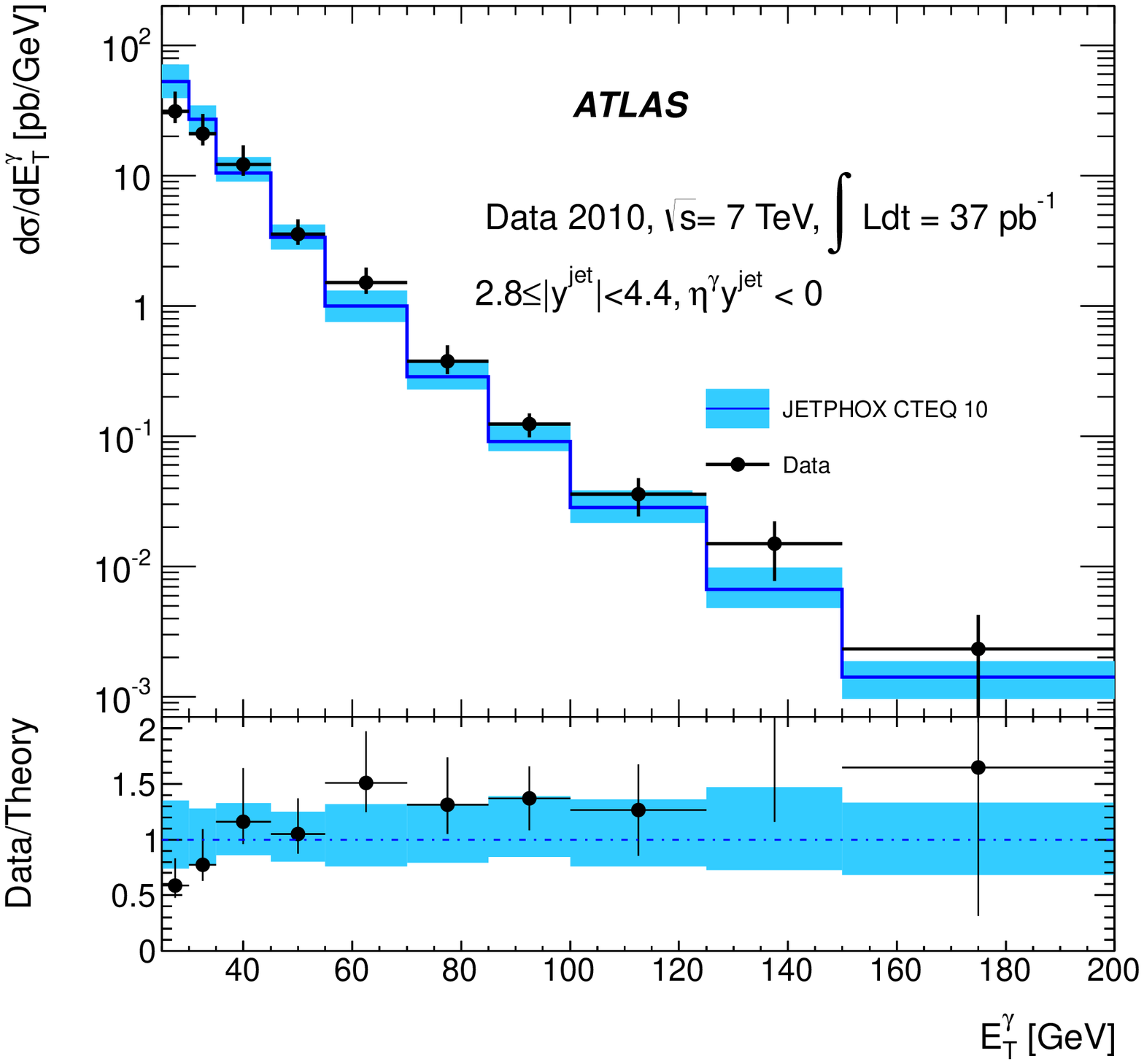}
    \caption{Top graphs: experimental (black dots) and theoretical (blue line) 
    photon + jet production cross sections, for the three same-sign (left column)
    and the three opposite-sign (right column) angular configurations. 
    The black error bars represent the total experimental uncertainty.
    The blue bands show the total 
    uncertainties on the theoretical predictions obtained with {\tt JETPHOX}. 
    Bottom graphs: ratio between the 
    measured and the predicted cross sections. 
    The blue bands show the theoretical uncertainties while the error bars show the experimental uncertainties on the ratio.
    First row: $|y^{\rm jet}|<1.2$.
    Second row: $1.2\le|y^{\rm jet}|<2.8$.
    Third row: $2.8\le|y^{\rm jet}|<4.4$.}
  \label{fig:datamc_comparison}
\end{figure*}

The measured $\ET^\gamma$-differential cross sections in the six
photon-jet angular configurations under study are shown, with the
theoretical cross sections overlaid, in 
Fig.~\ref{fig:datamc_comparison}.
The ratio between data and theory is also plotted, showing the relative
deviation of the measured cross section from the predicted cross 
section across the full $\ET^\gamma$ range on a linear scale. 
The error bars represent the combination of statistical and systematic
uncertainties, but are dominated by systematic uncertainties in all
regions. 
The numerical results are presented in Appendix~\ref{app:xsmeasured}.

The NLO pQCD predictions provided by {\tt JETPHOX} are in fair
agreement with the measured cross sections considering the given
experimental and theoretical systematic uncertainties. 
As already observed in previous measurements of the inclusive 
prompt photon cross section at the LHC~\cite{promptphotonpaper_2010,
promptphoton_confnote_2011,CMS_promptphotonpaper},
the data are consistently
lower than the theoretical prediction in the 
$\ET^{\gamma}$ $<45$~GeV
region, possibly suggesting an inaccuracy at low $\ET^{\gamma}$
of the NLO predictions and the need to perform the theoretical 
calculations at a higher order in perturbation theory.

\section{Conclusion}
\label{sec:Conclusion}
A measurement of the production cross section of an
isolated prompt photon in association with jets in $pp$ collisions at
a center-of-mass energy $\sqrt{s} = 7\TeV$ is presented.
The measurement uses an integrated luminosity of 37~\ipb~and
covers the region $x\gtrsim 0.001$ and 625~GeV$^2\leq Q^2 \leq
1.6 \times 10^5$ GeV$^2$, 
thus extending into kinematic regions previously unexplored
with this final state at either hadron or electron-proton colliders.
The differential cross section $d\sigma/d\ET^\gamma$, as a function of
the photon transverse energy, has been determined for isolated photons
in the pseudorapidity range $|\eta^\gamma|<1.37$ and transverse energy
$\ET^{\gamma}>25\GeV$, after integration over the jet transverse momenta
for $\pt^{\rm jet} > 20 \GeV$. A minimum separation of $\Delta R>1.0$ in the
$\{\eta,\phi\}$ plane is required between the leading jet and
the photon.
The cross sections are presented separately
for the three jet rapidity intervals $|y^{\rm jet}| < 1.2$, 
$1.2 \leq |y^{\rm jet}| < 2.8$ and $2.8 \leq |y^{\rm jet}| < 4.4$,
distinguishing between the same-sign
$(\eta^\gamma y^{\rm jet}\ge 0)$ and opposite-sign $(\eta^\gamma y^{\rm jet}< 0)$
configurations.
This subdivision allows the comparison between data and NLO
perturbative QCD predictions in configurations where the relative
contribution of the fragmentation component to the cross section and
the explored ranges of the incoming parton momentum fraction $x$ are different.
The NLO pQCD cross sections provided by {\tt JETPHOX} are in 
fair agreement with the measured ones considering the typical (10\% to 30\%)
experimental and theoretical systematic uncertainties.
In the $\ET^{\gamma}$ $<45$~GeV region, the NLO QCD calculation
consistently overestimates the measured cross section, as observed in previous
determinations of the inclusive prompt photon production cross section.

\section{Acknowledgements}

We thank CERN for the very successful operation of the LHC, as well as the
support staff from our institutions without whom ATLAS could not be
operated efficiently.

We acknowledge the support of ANPCyT, Argentina; YerPhI, Armenia; ARC,
Australia; BMWF, Austria; ANAS, Azerbaijan; SSTC, Belarus; CNPq and FAPESP,
Brazil; NSERC, NRC and CFI, Canada; CERN; CONICYT, Chile; CAS, MOST and NSFC,
China; COLCIENCIAS, Colombia; MSMT CR, MPO CR and VSC CR, Czech Republic;
DNRF, DNSRC and Lundbeck Foundation, Denmark; EPLANET and ERC, European Union;
IN2P3-CNRS, CEA-DSM/IRFU, France; GNAS, Georgia; BMBF, DFG, HGF, MPG and AvH
Foundation, Germany; GSRT, Greece; ISF, MINERVA, GIF, DIP and Benoziyo Center,
Israel; INFN, Italy; MEXT and JSPS, Japan; CNRST, Morocco; FOM and NWO,
Netherlands; RCN, Norway; MNiSW, Poland; GRICES and FCT, Portugal; MERYS
(MECTS), Romania; MES of Russia and ROSATOM, Russian Federation; JINR; MSTD,
Serbia; MSSR, Slovakia; ARRS and MVZT, Slovenia; DST/NRF, South Africa;
MICINN, Spain; SRC and Wallenberg Foundation, Sweden; SER, SNSF and Cantons of
Bern and Geneva, Switzerland; NSC, Taiwan; TAEK, Turkey; STFC, the Royal
Society and Leverhulme Trust, United Kingdom; DOE and NSF, United States of
America.

The crucial computing support from all WLCG partners is acknowledged
gratefully, in particular from CERN and the ATLAS Tier-1 facilities at
TRIUMF (Canada), NDGF (Denmark, Norway, Sweden), CC-IN2P3 (France),
KIT/GridKA (Germany), INFN-CNAF (Italy), NL-T1 (Netherlands), PIC (Spain),
ASGC (Taiwan), RAL (UK) and BNL (USA) and in the Tier-2 facilities
worldwide.

\bibliography{PhotonJetCrossSection}
\onecolumngrid
\clearpage

\appendix 

\onecolumngrid
\section{Theoretical photon + jet cross section}
\label{app:CStables_theory}
Tables~\ref{tab:jetphox__gc_jc_ss}-\ref{tab:jetphox__gc_jvf_os} show the theoretical photon + jet differential cross sections, in the six photon-jet angular configurations under study, computed as described in Sec.~\ref{sec:Theory}.

\begin{table}[!htb]
 \centering 
 \caption{NLO pQCD cross section prediction for the production of an isolated photon in the
   pseudorapidity range $0.00 \leq |\eta^\gamma| <
   1.37$ in association with a jet in the rapidity range $|y^{\rm jet}|<1.2$ and $\pT^{\rm
   jet}>$~20~GeV ($\eta^{\gamma}y^{\rm jet} \ge 0$). 
   The NLO pQCD cross section has been
   computed with {\tt JETPHOX} 1.3 using {\tt CT10} PDFs. 
   Details on the
   calculation  of the uncertainties are discussed in
   Sec.~\ref{sec:Theory}.  In the last column the nonperturbative
   correction factor that must multiply the {\tt JETPHOX} cross section is
   shown, with its uncertainty.} 
 \label{tab:jetphox__gc_jc_ss}
 \begin{tabular}{rrcccccc}
 \hline\hline 
$\ET^\gamma$~min & $\ET^\gamma$~max & \multicolumn{1}{c}{$\frac{d\sigma}{dE_{T}^{\gamma}}$} & stat     & scale    & PDF      & isolation & correction   \\
 & & & uncertainty & uncertainty & uncertainty& uncertainty & factor \\
$[{\rm GeV}]$ & $[{\rm GeV}]$ & $[{\rm pb/GeV}]$ & $[{\rm pb/GeV}]$ & $[{\rm pb/GeV}]$ & $[{\rm pb/GeV}]$ & $[{\rm pb/GeV}]$ & \\
\hline
25 & 30 & 550 & $\pm6$ & $^{+41}_{-41} $ & $^{+28}_{-36}$  &  $^{+9}_{-0} $ & 0.927$\pm$0.036\\
30 & 35 & 331 & $\pm3$ & $^{+51}_{-19} $ & $^{+16}_{-20}$  &  $^{+2}_{-0} $ & 0.951$\pm$0.034\\
35 & 45 & 156 & $\pm1$ & $^{+14}_{-13} $ & $^{+7}_{-8}$  &  $^{+0}_{-2} $ & 0.983$\pm$0.026\\
45 & 55 & 60.4 & $\pm0.4$ & $^{+6.6}_{-3.4} $ & $^{+2.7}_{-2.7}$  &  $^{+0.6}_{-0.0} $ & 0.992$\pm$0.020\\
55 & 70 & 24.2 & $\pm0.2$ & $^{+2.1}_{-1.6} $ & $^{+1.0}_{-0.9}$  &  $^{+0.0}_{-0.3} $ & 1.002$\pm$0.025\\
70 & 85 & 9.26 & $\pm0.06$ & $^{+0.92}_{-0.85} $ & $^{+0.34}_{-0.34}$  &  $^{+0.00}_{-0.14} $ & 0.995$\pm$0.026\\
85 & 100 & 4.21 & $\pm0.03$ & $^{+0.37}_{-0.34} $ & $^{+0.14}_{-0.14}$  &  $^{+0.01}_{-0.04} $ & 1.001$\pm$0.022\\
100 & 125 & 1.76 & $\pm0.01$ & $^{+0.16}_{-0.13} $ & $^{+0.05}_{-0.06}$  &  $^{+0.00}_{-0.01} $ & 0.996$\pm$0.017\\
125 & 150 & 0.699 & $\pm0.004$ & $^{+0.061}_{-0.056} $ & $^{+0.019}_{-0.020}$  &  $^{+0.005}_{-0.007} $ & 0.992$\pm$0.018\\
150 & 200 & 0.236 & $\pm0.001$ & $^{+0.024}_{-0.018} $ & $^{+0.006}_{-0.007}$  &  $^{+0.001}_{-0.003} $ & 0.997$\pm$0.016\\
200 & 400 & 0.0266 & $\pm0.0001$ & $^{+0.0026}_{-0.0023} $ & $^{+0.0008}_{-0.0008}$  &  $^{+0.0000}_{-0.0002} $ & 0.988$\pm$0.026\\
 \hline\hline 
 \end{tabular} 
 \end{table} 

\begin{table}[!htb]
 \centering 
 \caption{NLO pQCD cross section prediction for the production of an isolated photon
in the
   pseudorapidity range $0.00 \leq |\eta^\gamma| <
   1.37$ in association with a jet in the rapidity range $|y^{\rm
     jet}|<1.2$ and $\pT^{\rm
   jet}>$~20~GeV ($\eta^{\gamma}y^{\rm jet}<0$).
   The NLO pQCD cross section has been
   computed with {\tt JETPHOX} 1.3 using {\tt CT10} PDFs.
   Details on the
   calculation  of the uncertainties are discussed in
   Sec.~\ref{sec:Theory}.  In the last column the nonperturbative
   correction  factor that must multiply the {\tt JETPHOX} cross section is
   shown, with its uncertainty.}
 \label{tab:jetphox__gc_jc_os}
 \begin{tabular}{rrcccccc}
 \hline\hline 
$\ET^\gamma$~min & $\ET^\gamma$~max & \multicolumn{1}{c}{$\frac{d\sigma}{dE_{T}^{\gamma}}$} & stat     & scale    & PDF      & isolation & correction   \\
 & & & uncertainty & uncertainty & uncertainty& uncertainty & factor \\
$[{\rm GeV}]$ & $[{\rm GeV}]$ & $[{\rm pb/GeV}]$ & $[{\rm pb/GeV}]$ & $[{\rm pb/GeV}]$ & $[{\rm pb/GeV}]$ & $[{\rm pb/GeV}]$ & \\
\hline
25 & 30 & 420 & $\pm5$ & $^{+49}_{-40} $ & $^{+21}_{-29}$  &  $^{+8}_{-0} $ & 0.925$\pm$0.041\\
30 & 35 & 261 & $\pm2$ & $^{+27}_{-40} $ & $^{+11}_{-21}$  &  $^{+0}_{-7} $ & 0.943$\pm$0.049\\
35 & 45 & 118 & $\pm1$ & $^{+17}_{-16} $ & $^{+6}_{-6}$  &  $^{+2}_{-0} $ & 0.980$\pm$0.032\\
45 & 55 & 47.0 & $\pm0.3$ & $^{+4.3}_{-6.6} $ & $^{+2.6}_{-1.9}$  &  $^{+0.0}_{-0.7} $ & 0.979$\pm$0.029\\
55 & 70 & 17.2 & $\pm0.1$ & $^{+2.8}_{-1.6} $ & $^{+0.8}_{-0.7}$  &  $^{+0.0}_{-0.5} $ & 0.982$\pm$0.025\\
70 & 85 & 6.72 & $\pm0.05$ & $^{+0.62}_{-0.74} $ & $^{+0.28}_{-0.26}$  &  $^{+0.03}_{-0.02} $ & 0.995$\pm$0.018\\
85 & 100 & 2.93 & $\pm0.02$ & $^{+0.34}_{-0.25} $ & $^{+0.11}_{-0.11}$  &  $^{+0.03}_{-0.00} $ & 0.981$\pm$0.031\\
100 & 125 & 1.24 & $\pm0.01$ & $^{+0.14}_{-0.12} $ & $^{+0.04}_{-0.04}$  &  $^{+0.01}_{-0.02} $ & 0.989$\pm$0.025\\
125 & 150 & 0.469 & $\pm0.003$ & $^{+0.053}_{-0.039} $ & $^{+0.015}_{-0.016}$  &  $^{+0.005}_{-0.007} $ & 0.992$\pm$0.027\\
150 & 200 & 0.159 & $\pm0.001$ & $^{+0.020}_{-0.015} $ & $^{+0.005}_{-0.005}$  &  $^{+0.001}_{-0.001} $ & 0.984$\pm$0.019\\
200 & 400 & 0.0169 & $\pm0.0001$ & $^{+0.0017}_{-0.0015} $ & $^{+0.0007}_{-0.0007}$  &  $^{+0.0003}_{-0.0000} $ & 0.991$\pm$0.026\\
 \hline\hline 
 \end{tabular} 
 \end{table} 

\begin{table}[!htb]
 \centering 
 \caption{NLO pQCD cross section prediction for the production of an
   isolated photon in the pseudorapidity range $0.00 \leq |\eta^\gamma| <
   1.37$ in association with a jet in the rapidity range $1.2 \leq |y^{\rm
   jet}|<2.8$ and $\pT^{\rm jet}>$~20~GeV ($\eta^{\gamma}y^{\rm jet} \ge 
   0$). The NLO pQCD cross section has been computed with {\tt JETPHOX} 1.3
   using {\tt CT10} PDFs. Details on the calculation of the uncertainties
   are discussed in Sec.~\ref{sec:Theory}. In the last column the
   nonperturbative correction factor that must multiply the {\tt JETPHOX}
   cross section is shown, with its uncertainty.}
 \label{tab:jetphox__gc_jf_ss}
 \begin{tabular}{rrcccccc}
 \hline\hline 
$\ET^\gamma$~min & $\ET^\gamma$~max & \multicolumn{1}{c}{$\frac{d\sigma}{dE_{T}^{\gamma}}$} & stat     & scale    & PDF      & isolation & correction   \\
 & & & uncertainty & uncertainty & uncertainty& uncertainty & factor \\
$[{\rm GeV}]$ & $[{\rm GeV}]$ & $[{\rm pb/GeV}]$ & $[{\rm pb/GeV}]$ & $[{\rm pb/GeV}]$ & $[{\rm pb/GeV}]$ & $[{\rm pb/GeV}]$ & \\
\hline
25 & 30 & 434 & $\pm5$ & $^{+49}_{-35} $ & $^{+18}_{-23}$  &  $^{+3}_{-10 } $ & 0.925$\pm$0.054\\
30 & 35 & 258 & $\pm2$ & $^{+34}_{-30} $ & $^{+9}_{-12}$  &  $^{+0}_{-5 } $ & 0.955$\pm$0.034\\
35 & 45 & 120 & $\pm1$ & $^{+17}_{-14} $ & $^{+4}_{-5}$  &  $^{+0}_{-4 } $ & 0.988$\pm$0.021\\
45 & 55 & 42.4 & $\pm0.3$ & $^{+6.5}_{-3.7} $ & $^{+1.1}_{-1.5}$  &  $^{+0.9}_{-0.0 } $ & 0.993$\pm$0.023\\
55 & 70 & 16.7 & $\pm0.1$ & $^{+1.8}_{-2.0} $ & $^{+0.4}_{-0.5}$  &  $^{+0.5}_{-0.0 } $ & 0.997$\pm$0.025\\
70 & 85 & 6.02 & $\pm0.05$ & $^{+0.68}_{-0.55} $ & $^{+0.12}_{-0.17}$  &  $^{+0.05}_{-0.00 } $ & 0.994$\pm$0.015\\
85 & 100 & 2.66 & $\pm0.02$ & $^{+0.30}_{-0.24} $ & $^{+0.05}_{-0.07}$  &  $^{+0.02}_{-0.01 } $ & 0.989$\pm$0.020\\
100 & 125 & 1.09 & $\pm0.01$ & $^{+0.11}_{-0.10} $ & $^{+0.02}_{-0.03}$  &  $^{+0.01}_{-0.01 } $ & 0.997$\pm$0.020\\
125 & 150 & 0.401 & $\pm0.003$ & $^{+0.051}_{-0.035} $ & $^{+0.007}_{-0.009}$  &  $^{+0.002}_{-0.004 } $ & 0.998$\pm$0.020\\
150 & 200 & 0.125 & $\pm0.001$ & $^{+0.010}_{-0.011} $ & $^{+0.003}_{-0.003}$  &  $^{+0.000}_{-0.002 } $ & 0.994$\pm$0.023\\
200 & 400 & 0.0118 & $\pm0.0001$ & $^{+0.0011}_{-0.0012} $ & $^{+0.0003}_{-0.0003}$  &  $^{+0.0000}_{-0.0001 } $ & 0.993$\pm$0.017\\
 \hline\hline 
 \end{tabular} 
 \end{table} 

\begin{table}[!htb]
 \centering 
 \caption{NLO pQCD cross section prediction for the production of an
   isolated photon in the pseudorapidity range $0.00 \leq |\eta^\gamma| <
   1.37$ in association with a jet in the rapidity range $1.2 \leq |y^{\rm
   jet}|<2.8$ and $\pT^{\rm jet}>$~20~GeV ($\eta^{\gamma}y^{\rm jet} < 0$).
   The NLO pQCD cross section has been computed with {\tt JETPHOX} 1.3
   using {\tt CT10} PDFs. Details on the calculation of the uncertainties
   are discussed in Sec.~\ref{sec:Theory}. In the last column the
   nonperturbative correction factor that must multiply the {\tt JETPHOX}
   cross section is shown, with its uncertainty.}
 \label{tab:jetphox__gc_jf_os}
 \begin{tabular}{rrcccccc}
 \hline\hline 
$\ET^\gamma$~min & $\ET^\gamma$~max & \multicolumn{1}{c}{$\frac{d\sigma}{dE_{T}^{\gamma}}$} & stat     & scale    & PDF      & isolation & correction   \\
 & & & uncertainty & uncertainty & uncertainty& uncertainty & factor \\
$[{\rm GeV}]$ & $[{\rm GeV}]$ & $[{\rm pb/GeV}]$ & $[{\rm pb/GeV}]$ & $[{\rm pb/GeV}]$ & $[{\rm pb/GeV}]$ & $[{\rm pb/GeV}]$ & \\
\hline
25 & 30 & 260 & $\pm3$ & $^{+33}_{-44} $ & $^{+13}_{-12}$  &  $^{+0}_{-17 } $ & 0.935$\pm$0.075\\
30 & 35 & 141 & $\pm1$ & $^{+24}_{-23} $ & $^{+7}_{-6}$  &  $^{+0}_{-4 } $ & 0.909$\pm$0.055\\
35 & 45 & 60 & $\pm1$ & $^{+12}_{-9} $ & $^{+3}_{-2}$  &  $^{+1}_{-0} $ & 0.975$\pm$0.034\\
45 & 55 & 22.3 & $\pm0.2$ & $^{+3.8}_{-4.0} $ & $^{+0.8}_{-0.8}$  &  $^{+0.0}_{-1.0 } $ & 0.962$\pm$0.051\\
55 & 70 & 8.1 & $\pm0.1$ & $^{+1.5}_{-1.1} $ & $^{+0.3}_{-0.3}$  &  $^{+0.0}_{-0.3 } $ & 0.961$\pm$0.047\\
70 & 85 & 2.81 & $\pm0.02$ & $^{+0.40}_{-0.45} $ & $^{+0.09}_{-0.09}$  &  $^{+0.06}_{-0.07 } $ & 0.985$\pm$0.024\\
85 & 100 & 1.14 & $\pm0.01$ & $^{+0.24}_{-0.18} $ & $^{+0.04}_{-0.04}$  &  $^{+0.00}_{-0.02 } $ & 0.998$\pm$0.035\\
100 & 125 & 0.456 & $\pm0.004$ & $^{+0.078}_{-0.069} $ & $^{+0.016}_{-0.016}$  &  $^{+0.002}_{-0.012 } $ & 0.974$\pm$0.036\\
125 & 150 & 0.157 & $\pm0.002$ & $^{+0.040}_{-0.019} $ & $^{+0.006}_{-0.006}$  &  $^{+0.002}_{-0.000 } $ & 0.979$\pm$0.040\\
150 & 200 & 0.0481 & $\pm0.0005$ & $^{+0.0086}_{-0.0076} $ & $^{+0.0022}_{-0.0022}$  &  $^{+0.0010}_{-0.0000 } $ & 0.979$\pm$0.031\\
200 & 400 & 0.00422 & $\pm0.00005$ & $^{+0.00099}_{-0.00054} $ & $^{+0.00024}_{-0.00024}$  &  $^{+0.00002}_{-0.00005 } $ & 0.966$\pm$0.028\\
 \hline\hline 
 \end{tabular} 
 \end{table} 

\begin{table}[!htb]
 \centering 
 \caption{NLO pQCD cross section prediction for the production of an
   isolated photon in the pseudorapidity range $0.00 \leq |\eta^\gamma| <
   1.37$ in association with a jet in the rapidity range $2.8 \leq |y^{\rm
   jet}|<4.4$ and $\pT^{\rm jet}>$~20~GeV ($\eta^{\gamma}y^{\rm jet} \ge 
   0$). The NLO pQCD cross section has been computed with {\tt JETPHOX} 1.3
   using {\tt CT10} PDFs. Details on the calculation of the uncertainties
   are discussed in Sec.~\ref{sec:Theory}. In the last column the
   nonperturbative correction factor that must multiply the {\tt JETPHOX}
   cross section is shown, with its uncertainty.}
 \label{tab:jetphox__gc_jvf_ss}
 \begin{tabular}{rrcccccc}
 \hline\hline 
$\ET^\gamma$~min & $\ET^\gamma$~max & \multicolumn{1}{c}{$\frac{d\sigma}{dE_{T}^{\gamma}}$} & stat     & scale    & PDF      & isolation & correction   \\
 & & & uncertainty & uncertainty & uncertainty& uncertainty & factor \\
$[{\rm GeV}]$ & $[{\rm GeV}]$ & $[{\rm pb/GeV}]$ & $[{\rm pb/GeV}]$ & $[{\rm pb/GeV}]$ & $[{\rm pb/GeV}]$ & $[{\rm pb/GeV}]$ & \\
\hline
25 & 30 & 91 & $\pm2$ & $^{+18}_{-5} $ & $^{+2}_{-3}$  &  $^{+10}_{-0 } $ & 0.904$\pm$0.062\\
30 & 35 & 47 & $\pm1$ & $^{+12}_{-4} $ & $^{+1}_{-2}$  &  $^{+5}_{-0 } $ & 0.919$\pm$0.071\\
35 & 45 & 19.8 & $\pm0.3$ & $^{+4.1}_{-1.3} $ & $^{+0.4}_{-0.7}$  &  $^{+1.4}_{-0.0 } $ & 0.959$\pm$0.035\\
45 & 55 & 6.14 & $\pm0.11$ & $^{+2.31}_{-0.82} $ & $^{+0.18}_{-0.23}$  &  $^{+0.59}_{-0.00 } $ & 0.950$\pm$0.068\\
55 & 70 & 1.97 & $\pm0.04$ & $^{+0.38}_{-0.22} $ & $^{+0.07}_{-0.09}$  &  $^{+0.09}_{-0.04 } $ & 0.960$\pm$0.066\\
70 & 85 & 0.556 & $\pm0.013$ & $^{+0.147}_{-0.051} $ & $^{+0.026}_{-0.024}$  &  $^{+0.009}_{-0.002 } $ & 0.975$\pm$0.067\\
85 & 100 & 0.204 & $\pm0.005$ & $^{+0.049}_{-0.022} $ & $^{+0.012}_{-0.009}$  &  $^{+0.010}_{-0.003 } $ & 0.973$\pm$0.079\\
100 & 125 & 0.064 & $\pm0.002$ & $^{+0.008}_{-0.011} $ & $^{+0.004}_{-0.003}$  &  $^{+0.003}_{-0.000 } $ & 0.973$\pm$0.056\\
125 & 150 & 0.0146 & $\pm0.0005$ & $^{+0.0019}_{-0.0017} $ & $^{+0.0014}_{-0.0008}$  &  $^{+0.0012}_{-0.0004 } $ & 0.979$\pm$0.068\\
150 & 200 & 0.0027 & $\pm0.0001$ & $^{+0.0007}_{-0.0005} $ & $^{+0.0004}_{-0.0002}$  &  $^{+0.0004}_{-0.0000 } $ & 1.004$\pm$0.056\\
 \hline\hline 
 \end{tabular} 
 \end{table} 

\begin{table}[!htb]
 \centering 
 \caption{NLO pQCD cross section prediction for the production of an
   isolated photon in the pseudorapidity range $0.00 \leq |\eta^\gamma| <
   1.37$ in association with a jet in the rapidity range $2.8 \leq |y^{\rm
   jet}|<4.4$ and $\pT^{\rm jet}>$~20~GeV ($\eta^{\gamma}y^{\rm jet} < 0$).
   The NLO pQCD cross section has been computed with {\tt JETPHOX} 1.3
   using {\tt CT10} PDFs. Details on the calculation of the uncertainties
   are discussed in Sec.~\ref{sec:Theory}. In the last column the
   nonperturbative correction factor that must multiply the {\tt JETPHOX}
   cross section is shown, with its uncertainty.}
 \label{tab:jetphox__gc_jvf_os}
 \begin{tabular}{rrcccccc}
 \hline\hline 
$\ET^\gamma$~min & $\ET^\gamma$~max & \multicolumn{1}{c}{$\frac{d\sigma}{dE_{T}^{\gamma}}$} & stat     & scale    & PDF      & isolation & correction   \\
 & & & uncertainty & uncertainty & uncertainty& uncertainty & factor \\
$[{\rm GeV}]$ & $[{\rm GeV}]$ & $[{\rm pb/GeV}]$ & $[{\rm pb/GeV}]$ & $[{\rm pb/GeV}]$ & $[{\rm pb/GeV}]$ & $[{\rm pb/GeV}]$ & \\
\hline
25 & 30 & 53 & $\pm1$ & $^{+17}_{-8} $ & $^{+1}_{-1}$  &  $^{+0}_{-1 } $ & 0.84$\pm$0.26\\
30 & 35 & 27 & $\pm0$ & $^{+7}_{-5} $ & $^{+1}_{-1}$  &  $^{+1}_{-0 } $ & 0.81$\pm$0.21\\
35 & 45 & 10.4 & $\pm0.2$ & $^{+3.5}_{-1.2} $ & $^{+0.3}_{-0.3}$  &  $^{+0.5}_{-0.0 } $ & 0.92$\pm$0.09\\
45 & 55 & 3.37 & $\pm0.05$ & $^{+0.88}_{-0.69} $ & $^{+0.12}_{-0.12}$  &  $^{+0.23}_{-0.00 } $ & 0.88$\pm$0.08\\
55 & 70 & 1.00 & $\pm0.02$ & $^{+0.30}_{-0.21} $ & $^{+0.04}_{-0.04}$  &  $^{+0.10}_{-0.00 } $ & 0.93$\pm$0.15\\
70 & 85 & 0.287 & $\pm0.005$ & $^{+0.094}_{-0.058} $ & $^{+0.017}_{-0.014}$  &  $^{+0.005}_{-0.002 } $ & 0.95$\pm$0.06\\
85 & 100 & 0.091 & $\pm0.002$ & $^{+0.035}_{-0.010} $ & $^{+0.007}_{-0.005}$  &  $^{+0.004}_{-0.000 } $ & 0.97$\pm$0.10\\
100 & 125 & 0.028 & $\pm0.001$ & $^{+0.010}_{-0.006} $ & $^{+0.003}_{-0.002}$  &  $^{+0.000}_{-0.001 } $ & 0.94$\pm$0.12\\
125 & 150 & 0.0067 & $\pm0.0002$ & $^{+0.0030}_{-0.0016} $ & $^{+0.0008}_{-0.0005}$  &  $^{+0.0000}_{-0.0000 } $ & 1.00$\pm$0.11\\
150 & 200 & 0.0014 & $\pm0.0001$ & $^{+0.0004}_{-0.0004} $ & $^{+0.0002}_{-0.0001}$  &  $^{+0.0000}_{-0.0001 } $ & 0.92$\pm$0.21\\
 \hline\hline 
 \end{tabular} 
 \end{table} 

\clearpage

\section{Measured photon + jet cross section}
\label{app:xsmeasured}
Tables~\ref{tab:_gc_jc_ss_measured_summary}-\ref{tab:_gc_jvf_os_measured_summary}
show the measured photon + jet differential cross sections, in the six
photon-jet angular configurations under study, and the comparison to
the theoretical predictions. 

\begin{table}[!htb]
 \centering 
 \caption{Measured cross section as a function of the photon transverse energy, $\ET^\gamma$, for $|\eta^{\gamma}|\le 1.37$, $|y^{\rm jet}|<1.2 $ and  $\eta^{\gamma}y^{\rm jet} \ge 0 $. The last two columns show the cross section predicted by {\tt JETPHOX} and multiplied by the corresponding nonperturbative correction factor, and its uncertainty.} 
 \label{tab:_gc_jc_ss_measured_summary}
 \begin{tabular}{cccccccc}
 \hline\hline 
 \multicolumn{2}{c}{} & \multicolumn{4}{c}{Measured cross section} & \multicolumn{2}{c}{Predicted cross section} \\  
\hline
$\ET^\gamma$~min       & $\ET^\gamma$~max       &
 \multicolumn{1}{c}{$\frac{d\sigma}{dE_{T}^{\gamma}}$} & stat &  syst &
 total exp. uncertainty &  \multicolumn{1}{c}{$\frac{d\sigma}{dE_{T}^{\gamma}}$} & total theory uncertainty  \\
$[{\rm GeV}]$ & $[{\rm GeV}]$ & $[{\rm pb/GeV}]$ & $[{\rm pb/GeV}]$ & $[{\rm pb/GeV}]$ & $[{\rm pb/GeV}]$ & $[{\rm pb/GeV}]$ & $[{\rm pb/GeV}]$ \\
\hline
25 & 30 & 394 & $\pm8 $ & $^{+ 74}_{-30} $  & $^{+ 74}_{-31} $ & 510 & $^{+ 51}_{-55} $  \\ 
30 & 35 & 258 & $\pm6 $ & $^{+ 49}_{-23} $  & $^{+ 50}_{-23} $ & 315 & $^{+ 52}_{-29} $  \\ 
35 & 45 & 137 & $\pm3 $ & $^{+ 27}_{-13} $  & $^{+ 27}_{-13} $ & 153 & $^{+ 16}_{-16} $  \\ 
45 & 55 & 60.9 & $\pm0.7 $ & $^{+ 7.0}_{-5.2} $  & $^{+ 7.1}_{-5.2} $ & 59.9 & $^{+ 7.2}_{-4.5} $  \\ 
55 & 70 & 24.8 & $\pm0.3 $ & $^{+ 3.0}_{-2.3} $  & $^{+ 3.0}_{-2.4} $ & 24.3 & $^{+ 2.5}_{-2.0} $  \\ 
70 & 85 & 9.51 & $\pm0.20 $ & $^{+ 1.22}_{-0.98} $  & $^{+ 1.24}_{-1.00} $ & 9.21 & $^{+ 1.01}_{-0.95} $  \\ 
85 & 100 & 4.40 & $\pm0.15 $ & $^{+ 0.55}_{-0.48} $  & $^{+ 0.57}_{-0.50} $ & 4.21 & $^{+ 0.41}_{-0.39} $  \\ 
100 & 125 & 1.77 & $\pm0.07 $ & $^{+ 0.23}_{-0.20} $  & $^{+ 0.24}_{-0.22} $ & 1.76 & $^{+ 0.17}_{-0.15} $  \\ 
125 & 150 & 0.698 & $\pm0.038 $ & $^{+ 0.096}_{-0.085} $  & $^{+ 0.103}_{-0.093} $ & 0.693 & $^{+ 0.065}_{-0.061} $  \\ 
150 & 200 & 0.226 & $\pm0.017 $ & $^{+ 0.032}_{-0.029} $  & $^{+ 0.036}_{-0.034} $ & 0.236 & $^{+ 0.025}_{-0.020} $  \\ 
200 & 400 & 0.0283 & $\pm0.0028 $ & $^{+ 0.0041}_{-0.0038} $  & $^{+ 0.0050}_{-0.0048} $ & 0.0263 & $^{+ 0.0027}_{-0.0025} $  \\ 
 \hline\hline 
 \end{tabular} 
 \end{table} 

\begin{table}[!htb]
 \centering 
 \caption{Measured cross section as a function of the photon transverse energy, $\ET^\gamma$, for $|\eta^{\gamma}|\le 1.37$, $|y^{j}|<1.2 $ and  $\eta^{\gamma}y^{\rm jet} < 0 $. The last two columns show the cross section predicted by {\tt JETPHOX} and multiplied by the corresponding nonperturbative correction factor, and its uncertainty.} 
 \label{tab:_gc_jc_os_measured_summary}
 \begin{tabular}{cccccccc}
 \hline\hline
 \multicolumn{2}{c}{} & \multicolumn{4}{c}{Measured cross section} &
 \multicolumn{2}{c}{Predicted cross section} \\
\hline
$\ET^\gamma$~min       & $\ET^\gamma$~max       &
 \multicolumn{1}{c}{$\frac{d\sigma}{dE_{T}^{\gamma}}$} & stat &  syst &
 total exp. uncertainty &
 \multicolumn{1}{c}{$\frac{d\sigma}{dE_{T}^{\gamma}}$} & total theory
 uncertainty  \\
$[{\rm GeV}]$ & $[{\rm GeV}]$ & $[{\rm pb/GeV}]$ & $[{\rm pb/GeV}]$ &
 $[{\rm pb/GeV}]$ & $[{\rm pb/GeV}]$ & $[{\rm pb/GeV}]$ & $[{\rm pb/GeV}]$
 \\
\hline
25 & 30 & 324 & $\pm7 $ & $^{+ 64}_{-29} $  & $^{+ 65}_{-30} $ & 389 & $^{+ 53}_{-49} $  \\ 
30 & 35 & 201 & $\pm5 $ & $^{+ 41}_{-20} $  & $^{+ 41}_{-20} $ & 246 & $^{+ 30}_{-45} $  \\ 
35 & 45 & 112 & $\pm3 $ & $^{+ 23}_{-12} $  & $^{+ 23}_{-12} $ & 116 & $^{+ 18}_{-17} $  \\ 
45 & 55 & 45.5 & $\pm0.5 $ & $^{+ 5.6}_{-3.9} $  & $^{+ 5.6}_{-3.9} $ & 46.0 & $^{+ 5.1}_{-6.9} $  \\ 
55 & 70 & 18.3 & $\pm0.3 $ & $^{+ 2.4}_{-1.7} $  & $^{+ 2.4}_{-1.8} $ & 16.9 & $^{+ 2.9}_{-1.8} $  \\ 
70 & 85 & 7.18 & $\pm0.18 $ & $^{+ 0.97}_{-0.74} $  & $^{+ 0.99}_{-0.76} $ & 6.68 & $^{+ 0.69}_{-0.79} $  \\ 
85 & 100 & 3.26 & $\pm0.14 $ & $^{+ 0.38}_{-0.36} $  & $^{+ 0.40}_{-0.38} $ & 2.87 & $^{+ 0.36}_{-0.28} $  \\ 
100 & 125 & 1.36 & $\pm0.05 $ & $^{+ 0.17}_{-0.16} $  & $^{+ 0.17}_{-0.17} $ & 1.22 & $^{+ 0.15}_{-0.13} $  \\ 
125 & 150 & 0.503 & $\pm0.037 $ & $^{+ 0.065}_{-0.062} $  & $^{+ 0.075}_{-0.072} $ & 0.466 & $^{+ 0.056}_{-0.044} $  \\ 
150 & 200 & 0.156 & $\pm0.014 $ & $^{+ 0.023}_{-0.020} $  & $^{+ 0.027}_{-0.025} $ & 0.156 & $^{+ 0.021}_{-0.016} $  \\ 
200 & 400 & 0.0182 & $\pm0.0022 $ & $^{+ 0.0028}_{-0.0025} $  & $^{+ 0.0035}_{-0.0033} $ & 0.0167 & $^{+ 0.0019}_{-0.0017} $  \\ 
 \hline\hline 
 \end{tabular} 
 \end{table} 

\begin{table}[!htb]
 \centering 
 \caption{Measured cross section as a function of the photon transverse energy, $\ET^\gamma$, for $|\eta^{\gamma}|\le 1.37$, $1.2 \leq |y^{\rm jet}|<2.8 $ and  $\eta^{\gamma}y^{\rm jet} \ge 0 $. The last two columns show the cross section predicted by {\tt JETPHOX} and multiplied by the corresponding nonperturbative correction factor, and its uncertainty.} 
 \label{tab:_gc_jf_ss_measured_summary}
 \begin{tabular}{cccccccc}
 \hline\hline
 \multicolumn{2}{c}{} & \multicolumn{4}{c}{Measured cross section} &
 \multicolumn{2}{c}{Predicted cross section} \\
\hline
$\ET^\gamma$~min       & $\ET^\gamma$~max       &
 \multicolumn{1}{c}{$\frac{d\sigma}{dE_{T}^{\gamma}}$} & stat &  syst &
 total exp. uncertainty &
 \multicolumn{1}{c}{$\frac{d\sigma}{dE_{T}^{\gamma}}$} & total theory
 uncertainty  \\
$[{\rm GeV}]$ & $[{\rm GeV}]$ & $[{\rm pb/GeV}]$ & $[{\rm pb/GeV}]$ &
 $[{\rm pb/GeV}]$ & $[{\rm pb/GeV}]$ & $[{\rm pb/GeV}]$ & $[{\rm pb/GeV}]$
 \\
\hline
25 & 30 & 316 & $\pm7 $ & $^{+ 54}_{-30} $  & $^{+ 55}_{-31} $ & 401 & $^{+ 53}_{-46} $  \\ 
30 & 35 & 210 & $\pm6 $ & $^{+ 37}_{-22} $  & $^{+ 37}_{-23} $ & 247 & $^{+ 34}_{-32} $  \\ 
35 & 45 & 105 & $\pm2 $ & $^{+ 19}_{-12} $  & $^{+ 19}_{-12} $ & 119 & $^{+ 17}_{-16} $  \\ 
45 & 55 & 43.6 & $\pm0.6 $ & $^{+ 5.0}_{-3.7} $  & $^{+ 5.1}_{-3.8} $ & 42.1 & $^{+ 6.7}_{-4.2} $  \\ 
55 & 70 & 17.5 & $\pm0.3 $ & $^{+ 2.1}_{-1.7} $  & $^{+ 2.2}_{-1.7} $ & 16.6 & $^{+ 2.0}_{-2.1} $  \\ 
70 & 85 & 6.39 & $\pm0.17 $ & $^{+ 0.82}_{-0.66} $  & $^{+ 0.84}_{-0.68} $ & 5.99 & $^{+ 0.70}_{-0.58} $  \\ 
85 & 100 & 2.71 & $\pm0.10 $ & $^{+ 0.35}_{-0.29} $  & $^{+ 0.36}_{-0.31} $ & 2.63 & $^{+ 0.30}_{-0.25} $  \\ 
100 & 125 & 1.27 & $\pm0.05 $ & $^{+ 0.17}_{-0.15} $  & $^{+ 0.18}_{-0.16} $ & 1.08 & $^{+ 0.12}_{-0.10} $  \\ 
125 & 150 & 0.441 & $\pm0.028 $ & $^{+ 0.062}_{-0.054} $  & $^{+ 0.068}_{-0.061} $ & 0.400 & $^{+ 0.052}_{-0.037} $  \\ 
150 & 200 & 0.102 & $\pm0.012 $ & $^{+ 0.015}_{-0.013} $  & $^{+ 0.019}_{-0.018} $ & 0.125 & $^{+ 0.010}_{-0.012} $  \\ 
200 & 400 & 0.0090 & $\pm0.0017 $ & $^{+ 0.0013}_{-0.0012} $  & $^{+ 0.0022}_{-0.0021} $ & 0.0117 & $^{+ 0.0011}_{-0.0012} $  \\ 
 \hline\hline 
 \end{tabular} 
 \end{table} 

\begin{table}[!htb]
 \centering 
 \caption{Measured cross section as a function of the photon transverse energy, $\ET^\gamma$, for $|\eta^{\gamma}|\le 1.37$, $1.2 \leq |y^{\rm jet}|<2.8 $ and  $\eta^{\gamma}y^{\rm jet} < 0 $. The last two columns show the cross section predicted by {\tt JETPHOX} and multiplied by the corresponding nonperturbative correction factor, and its uncertainty.} 
 \label{tab:_gc_jf_os_measured_summary}
 \begin{tabular}{cccccccc}
 \hline\hline
 \multicolumn{2}{c}{} & \multicolumn{4}{c}{Measured cross section} &
 \multicolumn{2}{c}{Predicted cross section} \\
\hline
$\ET^\gamma$~min       & $\ET^\gamma$~max       &
 \multicolumn{1}{c}{$\frac{d\sigma}{dE_{T}^{\gamma}}$} & stat &  syst &
 total exp. uncertainty &
 \multicolumn{1}{c}{$\frac{d\sigma}{dE_{T}^{\gamma}}$} & total theory
 uncertainty  \\
$[{\rm GeV}]$ & $[{\rm GeV}]$ & $[{\rm pb/GeV}]$ & $[{\rm pb/GeV}]$ &
 $[{\rm pb/GeV}]$ & $[{\rm pb/GeV}]$ & $[{\rm pb/GeV}]$ & $[{\rm pb/GeV}]$
 \\
\hline
25 & 30 & 188 & $\pm6 $ & $^{+ 35}_{-27} $  & $^{+ 36}_{-27} $ & 243 & $^{+ 38}_{-49} $  \\ 
30 & 35 & 115 & $\pm4 $ & $^{+ 22}_{-17} $  & $^{+ 23}_{-18} $ & 128 & $^{+ 24}_{-23} $  \\ 
35 & 45 & 58 & $\pm2 $ & $^{+ 11}_{~-9} $  & $^{+ 12}_{~-9} $ & 58 & $^{+ 12}_{~-9} $  \\ 
45 & 55 & 24.1 & $\pm0.5 $ & $^{+ 3.1}_{-2.1} $  & $^{+ 3.1}_{-2.2} $ & 21.5 & $^{+ 3.9}_{-4.2} $  \\ 
55 & 70 & 8.8 & $\pm0.2 $ & $^{+ 1.2}_{-0.9} $  & $^{+ 1.2}_{-0.9} $ & 7.8 & $^{+ 1.5}_{-1.2} $  \\ 
70 & 85 & 3.32 & $\pm0.11 $ & $^{+ 0.46}_{-0.35} $  & $^{+ 0.48}_{-0.37} $ & 2.76 & $^{+ 0.41}_{-0.46} $  \\ 
85 & 100 & 1.49 & $\pm0.09 $ & $^{+ 0.16}_{-0.16} $  & $^{+ 0.19}_{-0.18} $ & 1.14 & $^{+ 0.25}_{-0.19} $  \\ 
100 & 125 & 0.54 & $\pm0.04 $ & $^{+ 0.06}_{-0.06} $  & $^{+ 0.07}_{-0.07} $ & 0.44 & $^{+ 0.08}_{-0.07} $  \\ 
125 & 150 & 0.175 & $\pm0.022 $ & $^{+ 0.022}_{-0.022} $  & $^{+ 0.031}_{-0.031} $ & 0.154 & $^{+ 0.040}_{-0.021} $  \\ 
150 & 200 & 0.055 & $\pm0.008 $ & $^{+ 0.007}_{-0.008} $  & $^{+ 0.011}_{-0.011} $ & 0.047 & $^{+ 0.009}_{-0.008} $  \\ 
200 & 400 & 0.0041 & $\pm0.0010 $ & $^{+ 0.0006}_{-0.0006} $  & $^{+ 0.0011}_{-0.0012} $ & 0.0041 & $^{+ 0.0010}_{-0.0006} $  \\ 
 \hline\hline 
 \end{tabular} 
 \end{table} 

\begin{table}[!htb]
 \centering 
 \caption{Measured cross section as a function of the photon transverse energy, $\ET^\gamma$, for $|\eta^{\gamma}|\le 1.37$, $2.8 \leq |y^{\rm jet}|<4.4 $ and  $\eta^{\gamma}y^{\rm jet} \ge 0 $. The last two columns show the cross section predicted by {\tt JETPHOX} and multiplied by the corresponding nonperturbative correction factor, and its uncertainty.} 
 \label{tab:_gc_jvf_ss_measured_summary}
 \begin{tabular}{cccccccc}
 \hline\hline
 \multicolumn{2}{c}{} & \multicolumn{4}{c}{Measured cross section} &
 \multicolumn{2}{c}{Predicted cross section} \\
\hline
$\ET^\gamma$~min       & $\ET^\gamma$~max       &
 \multicolumn{1}{c}{$\frac{d\sigma}{dE_{T}^{\gamma}}$} & stat &  syst &
 total exp. uncertainty &
 \multicolumn{1}{c}{$\frac{d\sigma}{dE_{T}^{\gamma}}$} & total theory
 uncertainty  \\
$[{\rm GeV}]$ & $[{\rm GeV}]$ & $[{\rm pb/GeV}]$ & $[{\rm pb/GeV}]$ &
 $[{\rm pb/GeV}]$ & $[{\rm pb/GeV}]$ & $[{\rm pb/GeV}]$ & $[{\rm pb/GeV}]$
 \\
\hline
25 & 30 & 66 & $\pm4 $ & $^{+ 18}_{-9} $  & $^{+ 19}_{-10} $ & 82 & $^{+ 20}_{-8} $  \\ 
30 & 35 & 46 & $\pm3 $ & $^{+ 13}_{-7} $  & $^{+ 13}_{-8} $ & 43 & $^{+ 12}_{-5} $  \\ 
35 & 45 & 20 & $\pm1 $ & $^{+ 6}_{-3} $  & $^{+ 6}_{-3} $ & 19 & $^{+ 4}_{-2} $  \\ 
45 & 55 & 8.1 & $\pm0.3 $ & $^{+ 1.4}_{-0.8} $  & $^{+ 1.4}_{-0.8} $ & 5.8 & $^{+ 2.3}_{-0.9} $  \\ 
55 & 70 & 2.4 & $\pm0.1 $ & $^{+ 0.4}_{-0.2} $  & $^{+ 0.4}_{-0.3} $ & 1.9 & $^{+ 0.4}_{-0.3} $  \\ 
70 & 85 & 0.86 & $\pm0.06 $ & $^{+ 0.15}_{-0.10} $  & $^{+ 0.17}_{-0.11} $ & 0.54 & $^{+ 0.15}_{-0.07} $  \\ 
85 & 100 & 0.24 & $\pm0.03 $ & $^{+ 0.03}_{-0.03} $  & $^{+ 0.04}_{-0.04} $ & 0.20 & $^{+ 0.05}_{-0.03} $  \\ 
100 & 125 & 0.07 & $\pm0.01 $ & $^{+ 0.01}_{-0.01} $  & $^{+ 0.02}_{-0.02} $ & 0.06 & $^{+ 0.01}_{-0.01} $  \\ 
125 & 150 & 0.014 & $\pm0.007 $ & $^{+ 0.002}_{-0.002} $  & $^{+ 0.007}_{-0.007} $ & 0.014 & $^{+ 0.003}_{-0.002} $  \\ 
150 & 200 & 0.0028 & $\pm0.0019 $ & $^{+ 0.0004}_{-0.0004} $  & $^{+ 0.0019}_{-0.0020} $ & 0.0027 & $^{+ 0.0009}_{-0.0006} $  \\ 
 \hline\hline 
 \end{tabular} 
 \end{table} 

\begin{table}[!htb]
 \centering 
 \caption{Measured cross section as a function of the photon transverse energy, $\ET^\gamma$, for $|\eta^{\gamma}|\le 1.37$, $2.8 \leq |y^{\rm jet}|<4.4 $ and  $\eta^{\gamma}y^{\rm jet} < 0 $. The last two columns show the cross section predicted by {\tt JETPHOX} and multiplied by the corresponding nonperturbative correction factor, and its uncertainty.} 
 \label{tab:_gc_jvf_os_measured_summary}
 \begin{tabular}{cccccccc}
 \hline\hline
 \multicolumn{2}{c}{} & \multicolumn{4}{c}{Measured cross section} &
 \multicolumn{2}{c}{Predicted cross section} \\
\hline
$\ET^\gamma$~min       & $\ET^\gamma$~max       &
 \multicolumn{1}{c}{$\frac{d\sigma}{dE_{T}^{\gamma}}$} & stat &  syst &
 total exp. uncertainty &
 \multicolumn{1}{c}{$\frac{d\sigma}{dE_{T}^{\gamma}}$} & total theory
 uncertainty  \\
$[{\rm GeV}]$ & $[{\rm GeV}]$ & $[{\rm pb/GeV}]$ & $[{\rm pb/GeV}]$ &
 $[{\rm pb/GeV}]$ & $[{\rm pb/GeV}]$ & $[{\rm pb/GeV}]$ & $[{\rm pb/GeV}]$
 \\
\hline
25 & 30 & 31 & $\pm4 $ & $^{+ 12}_{-4} $  & $^{+ 13}_{-6} $ & 44 & $^{+ 19}_{- 14} $  \\ 
30 & 35 & 21 & $\pm2 $ & $^{+ 8}_{-3} $  & $^{+ 9}_{-4} $ & 22 & $^{+ 8}_{-6} $  \\ 
35 & 45 & 12 & $\pm1 $ & $^{+ 5}_{-2} $  & $^{+ 5}_{-2} $ & 10 & $^{+ 3}_{-1} $  \\ 
45 & 55 & 3.5 & $\pm0.2 $ & $^{+ 1.1}_{-0.6} $  & $^{+ 1.1}_{-0.6} $ & 3.0 & $^{+ 0.8}_{-0.7} $  \\ 
55 & 70 & 1.5 & $\pm0.1 $ & $^{+ 0.5}_{-0.2} $  & $^{+ 0.5}_{-0.3} $ & 0.9 & $^{+ 0.3}_{-0.2} $  \\ 
70 & 85 & 0.38 & $\pm0.04 $ & $^{+ 0.11}_{-0.06} $  & $^{+ 0.12}_{-0.08} $ & 0.27 & $^{+ 0.09}_{-0.06} $  \\ 
85 & 100 & 0.12 & $\pm0.02 $ & $^{+ 0.01}_{-0.01} $  & $^{+ 0.03}_{-0.03} $ & 0.09 & $^{+ 0.04}_{-0.01} $  \\ 
100 & 125 & 0.036 & $\pm0.011 $ & $^{+ 0.002}_{-0.002} $  & $^{+ 0.011}_{-0.011} $ & 0.027 & $^{+ 0.010}_{-0.007} $  \\ 
125 & 150 & 0.015 & $\pm0.007 $ & $^{+ 0.002}_{-0.002} $  & $^{+ 0.007}_{-0.007} $ & 0.007 & $^{+ 0.003}_{-0.002} $  \\ 
150 & 200 & 0.0023 & $\pm0.0019 $ & $^{+ 0.0003}_{-0.0003} $  & $^{+ 0.0019}_{-0.0019} $ & 0.0013 & $^{+ 0.0005}_{-0.0005} $  \\ 
 \hline\hline 
 \end{tabular} 
 \end{table} 

\clearpage

\section{Experimental systematic uncertainties}
\label{app:systematics}
Tables~\ref{tab:geo_syst}-\ref{tab:FF_syst} show the experimental systematic 
uncertainties on the measured photon + jet differential cross sections,
in each $\ET^{\gamma}$ bin and photon-jet angular configuration under study,
for the various sources of sytematic uncertainties considered in Sec.~\ref{sec:xsection}.

\begin{table}[!htb]
 \centering 
 \caption{Relative systematic uncertainty (\%) introduced by the detector simulation. The $\ET^\gamma$~limits for the very forward jet configurations are given in parentheses.} 
 \label{tab:geo_syst}
 \begin{tabular}{ccccccc}
 \hline\hline 
 $\ET^\gamma$~range&$|y^{\rm jet}|<1.2$&$|y^{\rm jet}|<1.2$&$1.2\le|y^{\rm jet}|<2.8$&$1.2\le|y^{\rm jet}|<2.8$&$2.8\le|y^{\rm jet}|<4.4$&$2.8\le|y^{\rm jet}|<4.4$\\
$[{\rm GeV}]$& $\eta^{\gamma}y^{\rm jet}\ge0$ & $\eta^{\gamma}y^{\rm jet}<0$ & $\eta^{\gamma}y^{\rm jet}\ge0$ & $\eta^{\gamma}y^{\rm jet}<0$ & $\eta^{\gamma}y^{\rm jet}\ge0$ & $\eta^{\gamma}y^{\rm jet}<0$ \\
 \hline 
25-45 & $^{+8.4}_{-0.0} $ & $^{+7.0}_{-0.0} $ & $^{+5.6}_{-0.0} $ & $^{+5.8}_{-0.0} $ & $^{+10.7}_{-0.0} $ & $^{+22.6}_{-0.0} $ \\ 
45-400(200) & $^{+0.9}_{-0.0} $ & $^{+1.8}_{-0.0} $ & $^{+4.7}_{-0.0} $ & $^{+1.6}_{-0.0} $ & $^{+10.7}_{-0.0} $ & $^{+22.6}_{-0.0} $ \\ 
 \hline\hline 
 \end{tabular} 
 \end{table} 

\begin{table}[!htb]
 \centering 
 \caption{Relative systematic uncertainty (\%) introduced by the prompt photon simulation. The $\ET^\gamma$~limits for the very forward jet configurations are given in parentheses.} 
 \label{tab:sim_syst}
 \begin{tabular}{ccccccc}
 \hline\hline 
 $\ET^\gamma$~range&$|y^{\rm jet}|<1.2$&$|y^{\rm jet}|<1.2$&$1.2\le|y^{\rm jet}|<2.8$&$1.2\le|y^{\rm jet}|<2.8$&$2.8\le|y^{\rm jet}|<4.4$&$2.8\le|y^{\rm jet}|<4.4$\\
$[{\rm GeV}]$& $\eta^{\gamma}y^{\rm jet}\ge0$ & $\eta^{\gamma}y^{\rm jet}<0$ & $\eta^{\gamma}y^{\rm jet}\ge0$ & $\eta^{\gamma}y^{\rm jet}<0$ & $\eta^{\gamma}y^{\rm jet}\ge0$ & $\eta^{\gamma}y^{\rm jet}<0$ \\
 \hline 
25-45 & $^{+13.8}_{-4.0} $ & $^{+15.5}_{-5.7} $ & $^{+10.8}_{-5.2} $ & $^{+9.2}_{-11.3} $ & $^{+2.1}_{-5.0} $ & $^{+14.3}_{-4.5} $ \\ 
45-85 & $^{+7.6}_{-2.1} $ & $^{+8.3}_{-1.7} $ & $^{+5.8}_{-1.5} $ & $^{+8.2}_{-2.3} $ & $^{+8.0}_{-3.4} $ & $^{+15.5}_{-10.8} $ \\ 
85-150(200) & $^{+6.2}_{-0.6} $ & $^{+3.6}_{-1.3} $ & $^{+5.0}_{-0.8} $ & $^{+0.7}_{-0.4} $ & $^{+1.1}_{-2.5} $ & $^{+10.2}_{-7.9} $ \\ 
150-400 & $^{+5.1}_{-0.4} $ & $^{+6.9}_{-0.9} $ & $^{+3.6}_{-0.7} $ & $^{+2.4}_{-5.4} $ & n/a  & n/a  \\ 
 \hline\hline 
 \end{tabular} 
 \end{table} 

\begin{table}[!htb]
 \centering 
 \caption{Relative systematic uncertainty (\%) introduced by the electromagnetic energy scale uncertainty. The $\ET^\gamma$~limits for the very forward jet configurations are given in parentheses.} 
 \label{tab:ems_syst}
 \begin{tabular}{ccccccc}
 \hline\hline 
 $\ET^\gamma$~range&$|y^{\rm jet}|<1.2$&$|y^{\rm jet}|<1.2$&$1.2\le|y^{\rm jet}|<2.8$&$1.2\le|y^{\rm jet}|<2.8$&$2.8\le|y^{\rm jet}|<4.4$&$2.8\le|y^{\rm jet}|<4.4$\\
$[{\rm GeV}]$& $\eta^{\gamma}y^{\rm jet}\ge0$ & $\eta^{\gamma}y^{\rm jet}<0$ & $\eta^{\gamma}y^{\rm jet}\ge0$ & $\eta^{\gamma}y^{\rm jet}<0$ & $\eta^{\gamma}y^{\rm jet}\ge0$ & $\eta^{\gamma}y^{\rm jet}<0$ \\
 \hline 
25-45 & $^{+1.1}_{-0.6} $ & $^{+1.0}_{-0.6} $ & $^{+1.2}_{-0.4} $ & $^{+1.1}_{-0.5} $ & $^{+1.4}_{-0.7} $ & $^{+1.4}_{-1.0} $ \\ 
45-85 & $^{+2.2}_{-0.7} $ & $^{+2.7}_{-0.9} $ & $^{+2.0}_{-0.9} $ & $^{+3.1}_{-0.9} $ & $^{+3.7}_{-1.1} $ & $^{+3.7}_{-0.1} $ \\ 
85-150(200) & $^{+2.4}_{-0.6} $ & $^{+2.3}_{-1.1} $ & $^{+2.8}_{-1.0} $ & $^{+2.7}_{-1.4} $ & $^{+4.7}_{-2.7} $ & $^{+3.6}_{-1.5} $ \\ 
150-400 & $^{+2.9}_{-1.4} $ & $^{+2.5}_{-1.4} $ & $^{+2.8}_{-0.9} $ & $^{+3.7}_{-0.9} $ & n/a  & n/a  \\ 
 \hline\hline 
 \end{tabular} 
 \end{table} 

\begin{table}[!htb]
 \centering 
 \caption{Relative systematic uncertainty (\%) introduced by the jet energy scale uncertainty. The $\ET^\gamma$~limits for the very forward jet configurations are given in parentheses.} 
 \label{tab:jes_syst}
 \begin{tabular}{ccccccc}
 \hline\hline 
 $\ET^\gamma$~range&$|y^{\rm jet}|<1.2$&$|y^{\rm jet}|<1.2$&$1.2\le|y^{\rm jet}|<2.8$&$1.2\le|y^{\rm jet}|<2.8$&$2.8\le|y^{\rm jet}|<4.4$&$2.8\le|y^{\rm jet}|<4.4$\\
$[{\rm GeV}]$& $\eta^{\gamma}y^{\rm jet}\ge0$ & $\eta^{\gamma}y^{\rm jet}<0$ & $\eta^{\gamma}y^{\rm jet}\ge0$ & $\eta^{\gamma}y^{\rm jet}<0$ & $\eta^{\gamma}y^{\rm jet}\ge0$ & $\eta^{\gamma}y^{\rm jet}<0$ \\
 \hline 
25-45 & $^{+4.0}_{-3.6} $ & $^{+4.2}_{-3.8} $ & $^{+6.8}_{-5.4} $ & $^{+6.6}_{-5.9} $ & $^{+18.4}_{-9.4} $ & $^{+21.4}_{-9.6} $ \\ 
45-85 & $^{+0.2}_{-0.3} $ & $^{+0.3}_{-0.2} $ & $^{+0.9}_{-0.9} $ & $^{+1.5}_{-0.5} $ & $^{+5.2}_{-2.5} $ & $^{+7.6}_{-7.2} $ \\ 
85-150(200) & $^{+0.1}_{-0.1} $ & $^{+0.3}_{-0.3} $ & $^{+0.1}_{-0.0} $ & $^{+0.0}_{-0.1} $ & $^{+1.2}_{-1.5} $ & $^{+1.7}_{-0.0} $ \\ 
150-400 & $^{+0.0}_{-0.0} $ & $^{+0.0}_{-0.0} $ & $^{+0.0}_{-0.1} $ & $^{+0.1}_{-0.1} $ & n/a  & n/a  \\ 
 \hline\hline 
 \end{tabular} 
 \end{table} 

\begin{table}[!htb]
 \centering 
 \caption{Relative systematic uncertainty (\%) introduced by the electromagnetic energy resolution uncertainty. The $\ET^\gamma$~limits for the very forward jet configurations are given in parentheses.} 
 \label{tab:emr_syst}
 \begin{tabular}{ccccccc}
 \hline\hline 
 $\ET^\gamma$~range&$|y^{\rm jet}|<1.2$&$|y^{\rm jet}|<1.2$&$1.2\le|y^{\rm jet}|<2.8$&$1.2\le|y^{\rm jet}|<2.8$&$2.8\le|y^{\rm jet}|<4.4$&$2.8\le|y^{\rm jet}|<4.4$\\
$[{\rm GeV}]$& $\eta^{\gamma}y^{\rm jet}\ge0$ & $\eta^{\gamma}y^{\rm jet}<0$ & $\eta^{\gamma}y^{\rm jet}\ge0$ & $\eta^{\gamma}y^{\rm jet}<0$ & $\eta^{\gamma}y^{\rm jet}\ge0$ & $\eta^{\gamma}y^{\rm jet}<0$ \\
 \hline 
25-45 & $^{+0.1}_{-0.1} $ & $^{+0.1}_{-0.0} $ & $^{+0.0}_{-0.1} $ & $^{+0.3}_{-0.0} $ & $^{+0.0}_{-0.0} $ & $^{+0.4}_{-0.1} $ \\ 
45-85 & $^{+0.3}_{-0.2} $ & $^{+0.0}_{-0.2} $ & $^{+0.1}_{-0.2} $ & $^{+0.0}_{-0.6} $ & $^{+0.8}_{-0.7} $ & $^{+0.0}_{-0.8} $ \\ 
85-150(200) & $^{+0.0}_{-0.1} $ & $^{+0.3}_{-0.1} $ & $^{+0.3}_{-0.0} $ & $^{+0.1}_{-0.0} $ & $^{+0.1}_{-1.2} $ & $^{+0.6}_{-0.1} $ \\ 
150-400 & $^{+0.1}_{-0.0} $ & $^{+0.1}_{-0.0} $ & $^{+0.1}_{-0.1} $ & $^{+0.0}_{-0.4} $ & n/a  & n/a  \\ 
 \hline\hline 
 \end{tabular} 
 \end{table} 

\begin{table}[!htb]
 \centering 
 \caption{Relative systematic uncertainty (\%) introduced by the jet energy resolution uncertainty. The $\ET^\gamma$~limits for the very forward jet configurations are given in parentheses.} 
 \label{tab:jer_syst}
 \begin{tabular}{ccccccc}
 \hline\hline 
 $\ET^\gamma$~range&$|y^{\rm jet}|<1.2$&$|y^{\rm jet}|<1.2$&$1.2\le|y^{\rm jet}|<2.8$&$1.2\le|y^{\rm jet}|<2.8$&$2.8\le|y^{\rm jet}|<4.4$&$2.8\le|y^{\rm jet}|<4.4$\\
$[{\rm GeV}]$& $\eta^{\gamma}y^{\rm jet}\ge0$ & $\eta^{\gamma}y^{\rm jet}<0$ & $\eta^{\gamma}y^{\rm jet}\ge0$ & $\eta^{\gamma}y^{\rm jet}<0$ & $\eta^{\gamma}y^{\rm jet}\ge0$ & $\eta^{\gamma}y^{\rm jet}<0$ \\
 \hline 
25-45 & $^{+0.1}_{-0.0} $ & $^{+0.0}_{-0.0} $ & $^{+0.3}_{-0.0} $ & $^{+0.2}_{-0.0} $ & $^{+0.0}_{-0.2} $ & $^{+0.6}_{-0.0} $ \\ 
45-85 & $^{+0.0}_{-0.1} $ & $^{+0.1}_{-0.0} $ & $^{+0.0}_{-0.3} $ & $^{+0.7}_{-0.0} $ & $^{+0.7}_{-0.0} $ & $^{+0.0}_{-0.7} $ \\ 
85-150(200) & $^{+0.0}_{-0.0} $ & $^{+0.1}_{-0.0} $ & $^{+0.1}_{-0.0} $ & $^{+0.0}_{-0.5} $ & $^{+0.1}_{-0.0} $ & $^{+1.5}_{-0.0} $ \\ 
150-400 & $^{+0.0}_{-0.0} $ & $^{+0.0}_{-0.0} $ & $^{+0.0}_{-0.0} $ & $^{+0.2}_{-0.0} $ & n/a  & n/a  \\ 
 \hline\hline 
 \end{tabular} 
 \end{table} 

\begin{table}[!htb]
 \centering 
 \caption{Relative systematic uncertainty (\%) introduced by the background correlation. The $\ET^\gamma$~limits for the very forward jet configurations are given in parentheses.} 
 \label{tab:cor_syst}
 \begin{tabular}{ccccccc}
 \hline\hline 
 $\ET^\gamma$~range&$|y^{\rm jet}|<1.2$&$|y^{\rm jet}|<1.2$&$1.2\le|y^{\rm jet}|<2.8$&$1.2\le|y^{\rm jet}|<2.8$&$2.8\le|y^{\rm jet}|<4.4$&$2.8\le|y^{\rm jet}|<4.4$\\
$[{\rm GeV}]$& $\eta^{\gamma}y^{\rm jet}\ge0$ & $\eta^{\gamma}y^{\rm jet}<0$ & $\eta^{\gamma}y^{\rm jet}\ge0$ & $\eta^{\gamma}y^{\rm jet}<0$ & $\eta^{\gamma}y^{\rm jet}\ge0$ & $\eta^{\gamma}y^{\rm jet}<0$ \\
 \hline 
25-45 & $\pm$1.6 & $\pm$1.9 & $\pm$2.0 & $\pm$2.8 & $\pm$5.4 & $\pm$6.3 \\ 
45-85 & $\pm$0.7 & $\pm$0.6 & $\pm$0.7 & $\pm$1.1 & $\pm$1.6 & $\pm$3.4 \\ 
85-150(200) & $\pm$0.3 & $\pm$0.4 & $\pm$0.2 & $\pm$0.5 & $\pm$0.2 & $\pm$0.8 \\ 
150-400 & $\pm$0.1 & $\pm$0.2 & $\pm$0.8 & $\pm$0.6 & n/a  & n/a  \\ 
 \hline\hline 
 \end{tabular} 
 \end{table} 

\begin{table}[!htb]
 \centering 
 \caption{Relative systematic uncertainty (\%) introduced by the tightness control region in the purity extraction method. The $\ET^\gamma$~limits for the very forward jet configurations are given in parentheses.} 
 \label{tab:tig_syst}
 \begin{tabular}{ccccccc}
 \hline\hline 
 $\ET^\gamma$~range&$|y^{\rm jet}|<1.2$&$|y^{\rm jet}|<1.2$&$1.2\le|y^{\rm jet}|<2.8$&$1.2\le|y^{\rm jet}|<2.8$&$2.8\le|y^{\rm jet}|<4.4$&$2.8\le|y^{\rm jet}|<4.4$\\
$[{\rm GeV}]$& $\eta^{\gamma}y^{\rm jet}\ge0$ & $\eta^{\gamma}y^{\rm jet}<0$ & $\eta^{\gamma}y^{\rm jet}\ge0$ & $\eta^{\gamma}y^{\rm jet}<0$ & $\eta^{\gamma}y^{\rm jet}\ge0$ & $\eta^{\gamma}y^{\rm jet}<0$ \\
 \hline 
25-45 & $\pm$4.9 & $\pm$5.5 & $\pm$6.0 & $\pm$8.0 & $\pm$10.6 & $\pm$12.2 \\ 
45-85 & $\pm$1.3 & $\pm$2.0 & $\pm$2.1 & $\pm$3.0 & $\pm$2.4 & $\pm$3.3 \\ 
85-150(200) & $\pm$0.2 & $\pm$0.2 & $\pm$0.6 & $\pm$0.0 & $\pm$2.5 & $\pm$2.4 \\ 
150-400 & $\pm$0.9 & $\pm$0.5 & $\pm$1.2 & $\pm$1.4 & n/a  & n/a  \\ 
 \hline\hline 
 \end{tabular} 
 \end{table} 

\begin{table}[!htb]
 \centering 
 \caption{Relative systematic uncertainty (\%) introduced by the isolation control region in the purity extraction method. The $\ET^\gamma$~limits for the very forward jet configurations are given in parentheses.} 
 \label{tab:iso_syst}
 \begin{tabular}{ccccccc}
 \hline\hline 
 $\ET^\gamma$~range&$|y^{\rm jet}|<1.2$&$|y^{\rm jet}|<1.2$&$1.2\le|y^{\rm jet}|<2.8$&$1.2\le|y^{\rm jet}|<2.8$&$2.8\le|y^{\rm jet}|<4.4$&$2.8\le|y^{\rm jet}|<4.4$\\
$[{\rm GeV}]$& $\eta^{\gamma}y^{\rm jet}\ge0$ & $\eta^{\gamma}y^{\rm jet}<0$ & $\eta^{\gamma}y^{\rm jet}\ge0$ & $\eta^{\gamma}y^{\rm jet}<0$ & $\eta^{\gamma}y^{\rm jet}\ge0$ & $\eta^{\gamma}y^{\rm jet}<0$ \\
 \hline 
25-45 & $\pm$0.3 & $\pm$0.3 & $\pm$0.3 & $\pm$0.7 & $\pm$0.8 & $\pm$0.4 \\ 
45-85 & $\pm$0.3 & $\pm$0.3 & $\pm$0.4 & $\pm$0.3 & $\pm$0.7 & $\pm$0.3 \\ 
85-150(200) & $\pm$0.1 & $\pm$0.1 & $\pm$0.2 & $\pm$0.1 & $\pm$0.3 & $\pm$0.2 \\ 
150-400 & $\pm$0.1 & $\pm$0.3 & $\pm$0.4 & $\pm$0.1 & n/a  & n/a  \\ 
 \hline\hline 
 \end{tabular} 
 \end{table}

\begin{table}[!htb]
 \centering 
 \caption{Relative systematic uncertainty (\%) introduced by the shower shape corrections uncertainty. The $\ET^\gamma$~limits for the very forward jet configurations are given in parentheses.} 
 \label{tab:FF_syst}
 \begin{tabular}{ccccccc}
 \hline\hline 
 $\ET^\gamma$~range&$|y^{\rm jet}|<1.2$&$|y^{\rm jet}|<1.2$&$1.2\le|y^{\rm jet}|<2.8$&$1.2\le|y^{\rm jet}|<2.8$&$2.8\le|y^{\rm jet}|<4.4$&$2.8\le|y^{\rm jet}|<4.4$\\
$[{\rm GeV}]$& $\eta^{\gamma}y^{\rm jet}\ge0$ & $\eta^{\gamma}y^{\rm jet}<0$ & $\eta^{\gamma}y^{\rm jet}\ge0$ & $\eta^{\gamma}y^{\rm jet}<0$ & $\eta^{\gamma}y^{\rm jet}\ge0$ & $\eta^{\gamma}y^{\rm jet}<0$ \\
 \hline 
25-45 & $^{+2.9}_{-0.8} $ & $^{+2.6}_{-0.7} $ & $^{+3.5}_{-0.9} $ & $^{+3.9}_{-0.9} $ & $^{+3.3}_{-1.0} $ & $^{+2.5}_{-0.9} $ \\ 
45-85 & $^{+1.0}_{-0.3} $ & $^{+1.3}_{-0.4} $ & $^{+1.1}_{-0.3} $ & $^{+1.4}_{-0.4} $ & $^{+1.4}_{-0.5} $ & $^{+0.0}_{-2.0} $ \\ 
85-150(200) & $^{+0.2}_{-0.1} $ & $^{+0.0}_{-0.1} $ & $^{+0.3}_{-0.0} $ & $^{+0.0}_{-0.3} $ & $^{+0.0}_{-1.3} $ & $^{+0.8}_{-0.0} $ \\ 
150-400 & $^{+0.2}_{-0.1} $ & $^{+0.2}_{-0.0} $ & $^{+0.2}_{-0.1} $ & $^{+0.3}_{-0.0} $ & n/a  & n/a  \\ 
 \hline\hline 
 \end{tabular} 
 \end{table} 

\clearpage

\onecolumngrid
\begin{flushleft}
{\Large The ATLAS Collaboration}

\bigskip

G.~Aad$^{\rm 48}$,
B.~Abbott$^{\rm 110}$,
J.~Abdallah$^{\rm 11}$,
A.A.~Abdelalim$^{\rm 49}$,
A.~Abdesselam$^{\rm 117}$,
O.~Abdinov$^{\rm 10}$,
B.~Abi$^{\rm 111}$,
M.~Abolins$^{\rm 87}$,
O.S.~AbouZeid$^{\rm 157}$,
H.~Abramowicz$^{\rm 152}$
% ...
\bigskip

% ...
$^{10}$ Institute of Physics, Azerbaijan Academy of Sciences, Baku, Azerbaijan\\
$^{11}$ Institut de F\'isica d'Altes Energies and Departament de F\'isica de la Universitat Aut\`onoma  de Barcelona and ICREA, Barcelona, Spain\\
% ...
$^{48}$ Fakult\"{a}t f\"{u}r Mathematik und Physik, Albert-Ludwigs-Universit\"{a}t, Freiburg i.Br., Germany\\
$^{49}$ Section de Physique, Universit\'e de Gen\`eve, Geneva, Switzerland\\
% ...
$^{87}$ Department of Physics and Astronomy, Michigan State University, East Lansing MI, United States of America\\
% ...
$^{110}$ Homer L. Dodge Department of Physics and Astronomy, University of Oklahoma, Norman OK, United States of America\\
$^{111}$ Department of Physics, Oklahoma State University, Stillwater OK, United States of America\\
% ...
$^{117}$ Department of Physics, Oxford University, Oxford, United Kingdom\\
% ...
$^{152}$ Raymond and Beverly Sackler School of Physics and Astronomy, Tel Aviv University, Tel Aviv, Israel\\
% ...
$^{157}$ Department of Physics, University of Toronto, Toronto ON, Canada\\
% ...
\end{flushleft}


\end{document}